\documentclass[11pt,a4paper,authoryear,times]{article}
\usepackage{graphicx,float,latexsym,times}
\usepackage{amsfonts,amstext}

 \RequirePackage[OT1]{fontenc}
\long\def\symbolfootnote[#1]#2{\begingroup%
\def\thefootnote{\fnsymbol{footnote}}\footnote[#1]{#2}\endgroup}
\usepackage{amsmath}
\usepackage{amssymb}
\usepackage{theorem}
\usepackage{xspace}
\usepackage{geometry}
\usepackage{array}
\usepackage{subfigure}
\usepackage{epsfig}
\usepackage{hhline}
\usepackage{bbm}
\usepackage{natbib}
 \usepackage{multirow}
\usepackage{dcolumn}
\usepackage{color}
\usepackage{lscape}
\usepackage{hyperref}

\usepackage{colortbl, xcolor}
\definecolor{Gray}{gray}{0.9}

\theoremstyle{plain}% Theorem-like structures provided by amsthm.sty
\newtheorem{theorem}{Theorem}[section]
\newtheorem{lemma}[theorem]{Lemma}
\newtheorem{corollary}[theorem]{Corollary}

\theoremstyle{definition}

\newtheorem{remark}[theorem]{Remark}

\DeclareMathAccent{\widehat}{\mathord}{largesymbols}{"62}
\DeclareMathAccent{\widetilde}{\mathord}{largesymbols}{"65}

\def\eeX{\mathbb{X}}

\def\ebo{\textrm{\mathversion{bold}$\mathbf{\beta^0}$\mathversion{normal}}}

\def\eb{\textrm{\mathversion{bold}$\mathbf{\beta}$\mathversion{normal}}}

\def\e1{1\!\!1}

\theoremstyle{plain}% Theorem-like structures provided by amsthm.sty
\newcommand{\beqn}{\begin{eqnarray*}}
\newcommand{\eeqn}{\end{eqnarray*}}
  
\def\ee1{\textrm{\mathversion{bold}$\mathbf{\varepsilon}$\mathversion{normal}}}

\newcommand{\bfW}{{\bf W}}

\def\eu{\mathbf{{u}}}

\newcommand{\N}{\mathbb{N}}
\newcommand{\R}{\mathbb{R}}
\newcommand{\PP}{\mathbb{P}}

\def\eX{\mathbf{X}}
\def\eeX{\mathbb{X}}
\def\ex{\mathbf{x}}

\newcommand{\Var}{\mathbb{V}\mbox{ar}\,}
\newcommand{\E}{\mathbb{E}\,}

\def\argmin{\mathop{\mathrm{arg\,min}}} 

\begin{document}
%--------------------------------------------------
\title { Detection of similar  successive groups in a  model with diverging number of variable groups}
\date{}
\maketitle

%%%%%%%%% Authors, affiliations %%%%%%%%%%%%%%%%%%%%%%%%%%

\author{
\begin{center}
\vskip -1cm 
 Gabriela CIUPERCA$^1$,  Mat\'u\v{s} MACIAK$^{2}$, Fran\c cois WAHL$^1$\\
\small   {$^1$Institut Camille Jordan,  UMR 5208, Universit\'e Claude Bernard Lyon 1, France\\
 $^2$Charles University, Faculty of Mathematics and Physics, Prague, Czech Republic}
\end{center}
}
% \footnote{\noindent \small{CONTACT: Gabriela CIUPERCA,\\  Universit\'e de Lyon,Universit\'e Claude Bernard Lyon 1, CNRS, UMR 5208, Institut Camille Jordan, Bat.  Braconnier, M43, blvd du 11 novembre 1918, F - 69622 Villeurbanne Cedex, France\\
%	\textit{Email address}: Gabriela.Ciuperca@univ-lyon1.fr }}
\symbolfootnote[0]{\normalsize CONTACT:  G. Ciuperca,
Universit\'e de Lyon, Universit\'e Claude Bernard Lyon 1, CNRS, UMR 5208, Institut Camille Jordan, Bat.  Braconnier, M43, blvd du 11 novembre 1918, F - 69622 Villeurbanne Cedex, France; \textit{E-mail}: Gabriela.Ciuperca@univ-lyon1.fr}
 

%\ead{Gabriela.Ciuperca@univ-lyon1.fr, salloum@math.univ-lyon1.fr} 

\begin{abstract}
  In this paper, a linear model with grouped explanatory variables is considered. The idea is to perform an automatic detection of different successive groups of the unknown coefficients under the assumption that the number of groups is of the same order as the sample size. The standard least squares loss function and the quantile loss function are both used together with the fused and adaptive fused   penalty to simultaneously estimate and group the unknown parameters. The proper convergence rate is given for the obtained estimators and the upper bound for the number of different successive group is derived. A simulation study is used to compare the empirical performance of the proposed fused and adaptive fused estimators and a real application on the air quality data demonstrates the practical applicability of the proposed methods. 
\end{abstract}
%%%%%%%%%%%%%%%
%%%%%%%%%%%%%%%
  
\section{Introduction}
 The idea of this paper is to automatically detect  different  successive groups of unknown coefficients of some explanatory variables in a multivariate linear model. The number of groups is supposed to be of the same order as the number of observations. For a given loss function, the fused type penalties  allow this automatic detection of these successive groups of the unknown coefficients. Depending on the assumptions imposed on the model errors, two modeling frameworks are considered: either the standard least squares loss function is used or the robust quantile loss function is considered instead. Moreover, for each framework, two fused group  penalties are proposed: firstly, the fused-type penalty which is later used to construct the adaptive fused penalty leading to a more accurate  selection of different successive groups.
 For each of the two estimators    the convergence rates are provided and the upper bound for the number of the successive groups is derived.
  
 In order to highlight the novelty of our work, we firstly make the state of the art regarding the proposed fused method with the automatic detection of the grouped explanatory variables in the multivariate linear model. Let $g \in \mathbb{N}$ denote the number of   variable groups and let $n \in \mathbb{N}$ be the total number of the available observations. The fused quantile method for a particular case of non-grouped variables with the quantile level $\tau=0.5$ was already considered in \cite{Liu.Tao.18} where the fused LASSO penalized least absolute deviation (LAD) estimator in a high-dimensional linear model is discussed and the proper  convergence rate of the obtained estimator is derived together with  a linearized alternating directional method  for finding the numerical solution. 
 
The quantile linear model  with a finite number of non-grouped explanatory variables is investigated by \cite{Jiang-Wang-Bondell-13} and \cite{Jiang-Bondell-Wang-14} by utilizing the adaptive fused penalization. In \cite{Jiang-Wang-Bondell-13}, the oracle property for the difference in the estimated coefficients for two different quantile levels is proved. More precisely, an automatically detection of the unchanged quantile slope coefficients across various quantile levels is discussed. In  \cite{Jiang-Bondell-Wang-14}, the adapted fused method is used  to automatically select the explanatory variables and to identify their successive differences at the neighhoring quantile levels.  For a linear quantile regression with $g$ groups of explanatory variables, \cite{Ciuperca-17b} shows the oracle property for the adaptive fused estimator when $g=O(n^c)$, for $0 \leq c <1$.

  If the model errors satisfy the classical conditions (i.e., zero mean and bounded variance) then the least squares (LS) loss function is more appropriate: in such case,  the high-dimensional linear model with the automatic selection of the corresponding groups of the explanatory variables with the adaptive LASSO penalty is considered by \cite{Wei-Huang-10} for the Gaussian errors when the number of groups is much larger than the sample size ($g \gg n$) and by \cite{Zhang-Xiang-16} for non Gaussian errors.  
   These results are further elaborated in \cite{Wang-Tian-19} for a generalized linear model when $g=n^c$, with $0 < c < 1$. The automatic selection of the grouped variables is also considered in \cite{Guo.Zhang.Wang.Wu.15} where the SCAD penalty is utilized under the assumption that the number of groups can grow at a certain polynomial rate with $n$. A combination of the $L_1$ and $L_2$ norms under the Gaussian model errors is investigated in \cite{Campbell-Allen-17}, where the authors propose a structured variable selection in order to select at least one variable from each group.  To our best knowledge,   the only papers  considering the  fused penalty with the main focus on the selection of variable groups, there is a paper of \cite{Li.Mo.Yuan.Zhang.14} where the LS loss function is penalized with the fused LASSO penalty, where the $L_1$ norm si considered for the magnitudes of the parameters and also for the successive differences of between the estimated coefficients.
   
In the present paper, the penalty is of the fused type, that is, it is built against the $L_{q,1}$ norm (with $q \geq 1$) of the difference between two successive groups of parameters, while in the mentioned just before  papers,  the norm in the penalty is $L_{2,1}$ or $L_{1,1}$ of each parameter group, the goal being to automatically select the significant coefficient  groups and not the identical successive coefficient groups. In a model, it can have successive vectors of non-zero parameters that are not different. A practical example is given in Section \ref{numerical} of the present paper on the influence of the groups of meteorological variables measured every hour, on the daily maximum benzen concentration.   It is this type of automatic detection that interests us in the present work. Whether for the quantile  or LS methods, particular cases of $L_{q,1}$ penalties  of the difference between two successive groups of parameter vectors were considered within the  change-points automatic detection framework in linear model. Except that, in the literature, for linear models with change-points  the statistical model is different from that considered in this paper, because the parameter  number  of the model was constant. For the LS loss function, we have in \cite{Zhang.Geng.15} the sum of the  penalties $L_{2,1}$ and $L_{1,1}$, while in \cite{Qian.Su.16} the penalty is $L_{2,1}$. For the quantile loss function,  \cite{Ciuperca-Maciak-19} consider the $L_{2,1}$ penalty. \\
The paper is organized as follows. In Section \ref{Section_Model} we introduce the model, assumptions and general notation. In Section \ref{section_estimation}, fused and adaptive fused group estimators for LS and quantile loss function are defined and asymptotically studied. In Section \ref{numerical} we present a simulation study on the proposed estimators ans an application on real data. The proofs of the results in Section \ref{section_estimation} are given in Section \ref{Section_Proofs}.

\section{Model}
\label{Section_Model}
In this section we state the model definition and some general assumptions imposed on the model design. Let us start, however, with some notation which will be used throughout the paper:  All limits in  are taken with respect to $n \rightarrow \infty$; All vectors are columns and  matrices and vectors are denoted with a bold face; For some matrix $\boldsymbol{A}$ we denote its transpose as $\boldsymbol{A}^\top$ and for a set ${\cal A}$, we denote by $|{\cal A}|$ its cardinality and by $\overline{\cal A}$ its complement; Expressions $\mu_{\min}(.)$ and $\mu_{\max}(.)$ are used to refer to  the smallest and largest eigenvalue of some positive definite matrix and for  $\ex=(x_1, \cdots, x_p)^\top \in \R^p$ being some $p$ dimensional vector $\| \ex\|_q=\big(|x_1|^q+\cdots+|x_p|^q \big)^{1/q}$ denotes its $L_q$ norm while $\| \ex\|_\infty = \max(|x_1|, \cdots , |x_p|)$ stands for the maximum norm. If, in addition, $\ex=(\ex_1^\top, \cdots , \ex_g^\top)^\top$ is a vector split into $g$  subvectors, then $ \sum^g_{j=1}\|\ex_j\|_q$ defines for the  $L_{q,1}$  norm of $\ex$. 

Moreover,  $\eb_j = (\beta_{j,1}, \cdots , \beta_{j,p})^\top \in \mathbb{R}^p$ stands for the corresponding group specific vector of the dimension $p \in \mathbb{N}$, for any $j \in \{1, \dots, g\}$, where $g \in \mathbb{N}$ is the number of the successive groups. Last but not least,  $C$ denotes some positive generic constant not depending on $n$ which may take different values in different formulas throughout the paper.

 In the present paper, we consider a multivariate linear model with $g$ groups of explanatory  variables. The number of groups $g \in \mathbb{N}$   depends on the sample size $n \in \mathbb{N}$, $g$ being known, such that $g \leq n/p$, while the number of explanatory variables in each group is fixed and does not depend on $n$.  Without reducing any generality, it is assumed that each group of the explanatory variables contains the same number of variables, $p \in \mathbb{N}$. Thus, the overall number of all parameters in the regression model is $r_n \leq n$.

Let us consider the following linear model with the grouped explanatory variables
 \begin{equation}
 \label{M1}
 Y_i=\sum^g_{j=1}\eX_{i,j}^\top \eb_j+\varepsilon_i=\eeX_i^\top \eb^g+\varepsilon_i, \qquad i=1, \cdots, n,
 \end{equation}
 with $\eb^g \equiv (\eb_1^\top, \cdots , \eb_g^\top)^\top \in \R^{r_n}$, where $\eb_j \in \R^p$ is the vector of parameters for the group $j \in \{1, \dots, g\}$. For each observation $i \in \{1, \dots, n\}$, the vector  $\eeX_i \in \R^{r_n}$ contains the explanatory variables $\eX_{i,j} \in \mathbb{R}^p$ from all groups. These group specific explanatory variables  are assumed to be deterministic, for any $j=1, \dots, g$ and $i=1, \dots, n$. The error terms $\{\varepsilon_i\}_{i}$ are assumed to be independent and the response variable is denoted as $Y_i$., $\varepsilon_i$.
  The true (unknown) vector of parameters is $\ebo=(\eb^{0\top}_1, \dots , \eb^{0\top}_g)^\top$. For $p=1$, the model with ungrouped explanatory variables is obtained. Note that the order of appearance of the groups in the model in (\ref{M1}) is important and some natural ordering is required. 
  
  Given the data $\{(Y_i,\eeX_i^\top);~i = 1, \dots, n\}$ we would like to automatically  determine, using the fused method, whether two successive groups of the explanatory variables have the same influence on the response or not while, at the same time, quantifying the corresponding effect magnitudes. In addition to the example on the air pollution in Section \ref{numerical}, a nice demonstration of the practical applicability of the proposed estimation method can be also seen in the very recent work of \cite{Zhou-Liu-Nar-Ye.12}, where the fused group method allows for capturing the temporal smoothness of the predictive biomarkers on the cognitive scores in the progression of the  Alzheimer's disease. To achieve the sparsity property between two successive groups of the explanatory variables (in a sense that the corresponding vectors of estimated parameters for two successive groups are mostly the same), the fused and adaptive fused group estimators are proposed and studied with two loss functions: the standard least squares and the quantile check function. 
   
The asymptotic behavior of the group specific estimators for the fused and the adaptive  fused method with $n \geq g p$ are investigated for $n \to \infty$ where, in addition,  a deterministic sequence    $(b_n)_{n \in \N}$ is needed, such that
\begin{equation}
\label{abn}
b_n \rightarrow 0, \quad n^{1/2}b_n \rightarrow \infty.
\end{equation}
  
\noindent  \textbf{Example} \textit{of such sequence which satisfies (\ref{abn}) is $b_n=\big(n^{-1} \log n \big)^{1/2}$}. \\

Unlike  \cite{Ciuperca-17b}, where the number of groups is either fixed or it is of the order $n^c$, with $0 < c < 1$, the model in (\ref{M1}) assumes that the number of the successive groups may be of the same order as the sample size. A similar model is also considered in 
\cite{Ciuperca-Maciak-19}  where the change-point detection and estimation is performed in the quantile model with fused type penalty, however, for the unknown vector of parameters with the dimension $p$, not depending on $n$. The same model is also considered in \cite{Leonardi-Buhlmann.16} where the change-point locations are detected by utilizing the LS loss function  with the LASSO type penalty.

\bigskip
 
  \noindent\textbf{Assumptions}\\
    The following regularity assumptions imposed on the model design  are needed. The assumptions required for the model errors will be presented in Subsection \ref{Q_sect} for the quantile framework and in Subsection \ref{LS_sect} for the LS framework.
    
\begin{description}
\item \textbf{(A1)} $\max_{1 \leqslant i \leqslant n} \| \eeX_i\|_{\infty} \leq C_0$, for some constant $C_0 >0$.
 \item \textbf{(A2)} There exist two positive constants, $0 < m_0 \leq M_0 < \infty$, such that $$ m_0 \leq   \mu_{\min} ( n^{-1} \sum^{n}_{i=1} \eeX_{i} \eeX_{i}^\top) \leq   \mu_{\max} ( n^{-1} \sum^{n}_{i=1} \eeX_{i} \eeX_{i}^\top) \leq M_0.$$ 
\end{description}

Assumption (A1) is considered, for instance, in \cite{Leonardi-Buhlmann.16} for the high-dimensional regression model and, also,  by \cite{He.Kong.Wang.16} for the penalized quantile regression. Assumption (A2) is standard in the linear model to ensure the parameter identifiability (see, for example, \cite{Zhang-Xiang-16}, \cite{Ciuperca-17a}, \cite{Ciuperca-17b}, or \cite{Wu-Liu-09}).  

% A similar assumption to the one in (A2) is also considered in \cite{Fan-Li-Wang.17}, where the penalized Huber loss is used with random covariates $\ex$ of the dimension  $q$, such that $\log q=O(n^b)$, $0<b<1$, that is $$0 < \kappa_l \leq \mu_{min}\big(\E[\ex \ex^\top] \big) \leq \mu_{max}\big(\E[\ex \ex^\top] \big) \leq \kappa_u < \infty.$$  

\section{Estimation methods}
\label{section_estimation}
In this section, two estimation frameworks are presented: the automatic detection and estimation of the successive groups of the explanatory variables is considered under two different model error assumptions. For each framework, the asymptotic properties are investigated. Firstly, the fused group estimator is proposed and, afterwards, the adaptive version of the fused group estimator is defined. 

If the model errors  $\{\varepsilon_i\}_{i\leqslant i \leqslant n}$ in (\ref{M1}) do not meet the standard conditions for the existence of the first two moments then the robust version needs to be employed, therefore, the quantile estimation technique is appropriate. On the other hand, if the conditions $\E[\varepsilon_i]=0$ and $\Var[\varepsilon_i]< \infty$ are satisfied, the penalized LS method is  considered. The  main results are presented for both scenarios in next two subsections while the proofs are all postponed to Section \ref{Section_Proofs}.
\subsection{\textbf{\textit{Quantile loss function}}}
\label{Q_sect}
Let the model errors in \eqref{M1} satisfy the following:

\begin{description}
\item \textbf{(A3)} Random errors $\varepsilon_i$, for $i = 1, \dots, n$, are independent and identically distributed (i.i.d.) with the continuous distribution function $F$, such that $F(0) = \PP[\varepsilon  \leq 0]=\tau$, for some known $\tau \in (0,1)$. The corresponding density function $f$ with the nonzero compact support ${\cal G} \subseteq \R$ is supposed to be continuous and strictly positive in a neighborhood of zero. Moreover, the first derivative of $f$ is bounded in a neighborhood of zero.

\end{description}

Assumption (A3) on the errors  is standard for the quantile regression models when the number of parameters depends on the sample size $n \in \mathbb{N}$ (see, for instance, \cite{Ciuperca-17a} and  \cite{Wu-Liu-09}). The standard assumptions  $\E[\varepsilon]=0$ and $\E[\varepsilon^2]< \infty$ are not considered and, therefore, the least squares method is not appropriate. Since $\PP[\varepsilon <0]=\tau$, we can consider the quantile method with the fixed quantile level $\tau \in (0,1)$, with the corresponding  quantile check function  $\rho_\tau(u)=u(\tau-\e1_{\{u <0\}})$, for $u \in \mathbb{R}$.  Thus, for the model in (\ref{M1}) the following quantile random process is obtained
 \begin{equation}
 \label{Gn}
 G_n(\eb^g) \equiv \sum^n_{i=1} \rho_ \tau(Y_i -\eeX_i^\top \eb^g),
 \end{equation}
with  the group quantile estimator defined as
 \begin{equation}
 \label{qest}
 \widetilde{{\eb}^{g}}\equiv \argmin_{\eb^g \in \R^{r_n}}G_n(\eb^g). 
 \end{equation}
 For the particular case of $\tau= 0 .5$  we obtain the median regression, for which the quantile process and the associated estimator (\ref{qest}) are reduced to the absolute deviation process and the least absolute deviation estimator respectively. The following Lemma gives the appropriate convergence rate of the group quantile estimator $\widetilde{\eb^{g}}$. 
 
 \begin{lemma}
 \label{Lemma 1}
 Under Assumptions (A1), (A2), and (A3) it holds that 
 $$\|\widetilde{{\eb}^{g}} - \ebo\|_1=O_{\PP}(b_n), $$ 
 where $(b_n)_{n \in \N}$ is the sequence defined in \eqref{abn}.
 \end{lemma}
The convergence rate of the group quantile estimator for the number of groups  $g=O(n)$ is different from that obtained when $g=O(n^c)$, with $0 \leq c<1$. Indeed, for $0 \leq c < 1$ the convergence rate of $\widetilde{{\eb}^{g}} $ is of the order $O_\PP(g n^{-1})^{1/2}=O_\PP(n^{(c-1)/2})$ (see Lemma 1 of \cite{Ciuperca-17a}) and the convergence rate of $\widetilde{{\eb}^{g}}$ from \eqref{qest} can not be obtained as a straightforward extension of the situation where $g=O(n^c)$ for $0 \leq c<1$, when $c \to 1$.
 
In order to preserve the group effect of the explanatory variables and to simultaneously 
 detect the successive groups of identical parameter vectors the $L_{q,1}$ norm of the consecutive differences $\eb_j-\eb_{j-1}$, for $j=2, \cdots , g$, is used as a penalty with some  $q \geq 1$ fixed. Thus, the following quantile process is considered
 
 \begin{equation}
 \label{Qnb}
 Q_n(\eb^g) \equiv G_n( \eb^g)+n \lambda_n \sum^g_{j=2} \| \eb_j-\eb_{j-1}\|_q.
 \end{equation}
 
  For $q=1$ the relation is (\ref{Qnb}) gives the process penalized with  the standard $L_1$ norm while for $q=2$ the process is penalized by the $L_{2,1}$ norm.  The positive sequence $(\lambda_n)_{n \in \N}$ plays a role of a tuning parameter, such that it converges  to  zero as the sample size tends to infinity. An additional condition on $(\lambda_n)_{n \in\N}$ will be given later when formulating the  theorems with the main results. 
  
Based on the penalized process in \eqref{Qnb}, the corresponding fused group quantile estimator is obtained as 
 \begin{equation}
 \label{heag}
 \widehat{{\eb}^g}\equiv  \argmin_{\eb^g \in \R^{r_n}} Q_n(\eb^g),
 \end{equation}
 where $ \widehat{{\eb}^g}=\big( \widehat{\eb}_1^\top, \dots ,  \widehat{\eb}_g^\top  \big)^\top$. 
 The estimator $\widehat{{\eb}^g}$ depends on the norm considered in the penalty term of random process in \eqref{Qnb} and, also, the tuning parameter  $\lambda_n > 0$.
 
   Let us define the set of indexes which form the true different successive groups 
 \begin{equation}
 \label{dA0}
 {\cal B}^0=\big\{ j \in \{2, \cdots , g \}; \eb_j^0 \neq \eb^0_{j-1} \big\}.
 \end{equation}
Since the values of the true parameter vector $\ebo$ are unknown the set $   {\cal B}^0$  is left unknown too. Therefore, an analogous set is considered with respect to the differences of the estimated parameters of two successive groups as
 \[
  \widehat{\cal B}_n=\big\{ j \in \{2, \cdots , g \}; \widehat{\eb}_j \neq \widehat{\eb}_{j-1} \big\}.
 \]
 It is obvious, that this set is used to provide a reasonable estimate for ${\cal B}^0$.

\begin{remark}
\label{nouv_p1}
The results obtained in this section are also valid for $p=1$, which is, to the authors' best knowledge, the case which has not been previously considered with in any literature. The  number of the groups in \cite{Ciuperca-17b} is of order $n^c$, with $0 \leq c <1$ and, moreover, in \cite{Ciuperca-17b}, the goal is to select the groups of significant variables simultaneously with the group's inheritance. 
\end{remark}

 The following theorem provides the convergence rate of the  fused group  quantile estimator defined in \eqref{heag}, under the additional assumption that there is only a finite number of the successive groups with different coefficients. For a suitable choice of the tuning parameter this convergence rate is of the same order as the sequence $(b_n)$ and, moreover, it is the same as the one obtained for the group quantile estimator in Lemma \ref{Lemma 1}. The convergence rate of $\widehat{{\eb}^g}$ does not depend on the $L_q$ norm considered in the penalty term in  \eqref{Qnb}.
 \begin{theorem}
 \label{theorem v_conv}
Under Assumptions (A1), (A2), and (A3), the condition in \eqref{abn}, if, moreover, $|{\cal B}^0|< \infty$ and $\lambda_nb_n^{-1} {\underset{n \rightarrow \infty}{\longrightarrow}} 0$, then  
$$\|  \widehat{{\eb}^g} -\ebo\|_1=O_{\PP} (b_n).$$
 \end{theorem}

 \noindent \textbf{Examples} \textit{of such sequences $(\lambda_n)_{n \in \N}$, $(b_n)_{n \in \N}$ which satisfy \eqref{abn} and $\lambda_nb_n^{-1} {\underset{n \rightarrow \infty}{\longrightarrow}} 0$  are  $\lambda_n = n^{-1} (\log n)^{1/2}$ and $b_n=\left(n^{-1}\log n \right)^{1/2}$.}\\

 Similarly as for the standard LASSO type penalties, the consistent selection of the different successive groups does not occur with the probability converging to 1 and some overfitting is observed. The missclassification error $|  \widehat{\cal B}_n \setminus {\cal B}^0|$ is used to assess 
 the number of the different successive groups being mistakenly detected as different. The following theorem provides the upper bound for this missclassification error. 
 
 \begin{theorem}
 \label{Proposition cardA}
 Under the same assumptions as in Theorem \ref{theorem v_conv},   there exists a positive constant $C_1 > 0$, such that 
 \[ \lim_{n \rightarrow \infty} \PP\bigg[|    \widehat{\cal B}_n \setminus {\cal B}^0| \leq C_1 \max  \bigg( \frac{b_n }{ \lambda_n} , \frac{1}{b_n} \bigg)\bigg]=1.
 \]
 \end{theorem}

 Note, that the upper bound in Theorem \ref{Proposition cardA} depends on the tuning parameter $\lambda_n > 0$ and the sequence $(b_n)_{n \in \N}$ abd thus, it can be hypothetically unbounded from above. Nevertheless, this result provides the upper bound for the number of elements in $\widehat{\cal B}_n$, more specifically, it gives the upper bound for the number of successive groups  of explanatory variables which have different estimated effect on the response variable. 
 
 \begin{corollary} 
 Since  $|{\cal B}^0|< \infty$ and $\big|  \widehat{\cal B}_n \setminus {\cal B}^0 \big| \geq |\widehat{\cal B}_n | - |{\cal B}^0 |$ with probability one, we can deduce by Theorem \ref{Proposition cardA}, that 
 $$ \lim_{n \rightarrow \infty} \PP\big[|    \widehat{\cal B}_n | \leq C \max  \big( {b_n }\lambda_n^{-1} ,b_n^{-1} \big)\big]=1.$$
 \end{corollary}

\begin{remark}  
\label{R21}
For instance, if $\lambda_n=n^{-1}(\log n)^{5/2} $ and  $b_n=\left(n^{-1}\log n \right)^{1/2}$,  then the upper bound given by Theorem \ref{Proposition cardA} is
\[
|\widehat{\cal B}_n| \leq C \max \big( n^{1/2} (\log n)^{-2},  n^{1/2} (\log n)^{-1/2} \big)= C n^{1/2} (\log n)^{-1/2},
\]
 which implies that the number of elements contained by $ \widehat{\cal B}_n$ is much smaller than $n^{1/2}$, however, it can converge to infinity for $n \to \infty$.
 \end{remark}

To improve the estimation accuracy of ${\cal B}^0$  we consider an adaptive penalty constructed on the basis of the  estimator in \eqref{heag}. Let us consider the random process
 \begin{equation}
 \label{quantile_adfuslasso}
 \overset{\vee}{ Q}_n(\eb^g) \equiv G_n( \eb^g)+n \lambda_n \sum^g_{j=2} \widehat{\omega}_{n,j} \| \eb_j-\eb_{j-1}\|_q,
 \end{equation} 
 with the adaptive weights 
 \[\widehat{\omega}_{n,j} \equiv 1/\max \big(n^{-1/2}, \sum^p_{k=1}|\widehat{\beta}_{j,k} - \widehat{\beta}_{j-1,k}|^\gamma \big),
 \]
  for   a fixed constant $\gamma>0$, where  $\widehat{\eb}_j = (\widehat{\beta}_{j,1}^\top, \dots, \widehat{\beta}_{j,p}^\top)^\top$.  Let us remark that for $j \not \in \widehat{\cal B}_n$ we have $\widehat{\eb}_j - \widehat{\eb}_{j-1}=\textbf{0}_p$.  The tuning parameter sequences in relations (\ref{Qnb}) and (\ref{quantile_adfuslasso}) may be different, both with a convergence rate faster than the sequence $(b_n)_{n \in \N}$. Therefore, the choice of $n^{-1/2}$ in $ \widehat{\omega}_{n,j}$ is used as deterministic sequence that converges to 0 when $\widehat{\eb}_j = \widehat{\eb}_{j-1}$, however, with the rate faster than $b_n$ because of the condition $n^{1/2} b_n \rightarrow \infty$ in \eqref{abn}. Notice that $\overset{\vee}{ Q}_n(\ebo) \equiv G_n( \ebo)+n \lambda_n \sum^g_{j=2} \widehat{\omega}_{n,j} \| \eb^0_j-\eb^0_{j-1}\|_q$ and the adaptive fused group quantile estimator for $\ebo$ is defined as
  \begin{equation*}
 \overset{\vee}{{\eb}^g}\equiv \argmin_{\eb^g \in \R^{r_n}} \overset{\vee}{ Q}_n(\eb^g),
 \end{equation*}
where $\overset{\vee}{{\eb}^g}=\big(\overset{\vee}{\eb}_1^\top, \cdots , \overset{\vee}{\eb}_g^\top \big)^\top$.  By Theorem \ref{theorem v_conv}, we have that  for all $j \in {\cal B}^0$   there exists a constant $c>0$ such that 
\begin{equation}
\label{onj}
 \lim_{n \rightarrow \infty}\PP\big[\widehat{\omega}_{n,j} >c \; | \; j \in {\cal B}^0 \big] = 1.
\end{equation}
Therefore, taking into account the relation in \eqref{onj} and the fact that $\gamma > 0$ a  similar proof to that of Theorem \ref{theorem v_conv} can be used to derive the convergence rate of  $ \overset{\vee}{{\eb}^g}$ which is the same as for $\widehat{\eb^g}$.

 \begin{theorem}
 \label{theorem v_conv_bis}
Under Assumptions (A1), (A2), and (A3), the condition in \eqref{abn}, if  $|{\cal B}^0|< \infty$ the for any sequence $(\lambda_n)_{n \in \N}$ such that $\lambda_nb_n^{-1} {\underset{n \rightarrow \infty}{\longrightarrow}} 0$ it holds that
$$\|\overset{\vee}{{\eb}^g} -\ebo\|_1=O_{\PP} (b_n).$$
 \end{theorem}  
 
Considering the adaptive fused group quantile  estimator $\overset{\vee}{{\eb}^g}$ we can also define an updated estimator for the set ${\cal B}^0$ as
   \[
  \overset{\vee}{{\cal B}_n} \equiv \big\{ j \in \{2, \cdots , g \}; \overset{\vee}{\eb}_j \neq \overset{\vee}{\eb}_{j-1} \big\},
 \]
 which is indeed more appropriate as shown by the next theorem where the upper bound for the cardinality of  $ \overset{\vee}{{\cal B}_n}\setminus {\cal B}^0$ is proved to be  much smaller than the one for $ \widehat{\cal B}_n \setminus {\cal B}^0$ in Theorem \ref{Proposition cardA}.
 \begin{theorem}
 \label{Proposition cardA_bis}
 Under the same assumptions as in  Theorem \ref{theorem v_conv_bis},   there exist a positive constant $C_2$ such that, 
 \[ \lim_{n \rightarrow \infty} \PP\bigg[|   \overset{\vee}{{\cal B}_n}\setminus {\cal B}^0| \leq C_2 \max \big(n^{-1/2}, b^\gamma_n \big) \max  \bigg( \frac{b_n }{ \lambda_n} , \frac{1}{b_n} \bigg)\bigg]=1.
 \]
 \end{theorem}

\begin{remark} 
\label{R22}
(i)  For $\gamma >1$ and the tuning parameter $(\lambda_n)_{n \in \N}$ such that $n^{-1/2}b_n \lambda_n^{-1} \rightarrow 0$ and $b^{\gamma+1}_n \lambda_n^{-1} \rightarrow0$, we obtain that $ \max \big(n^{-1/2}, b^\gamma_n \big) \max  \big( b_n \lambda_n^{-1} ,  b_n^{-1} \big) \rightarrow 0$, as $n \rightarrow \infty$. The examples of sequences $(\lambda_n)$ and $(b_n)$ from Remark \ref{R21} satisfy these conditions. \\
 (ii) If $0 <\gamma \leq 1$ then, $\max \big(n^{-1/2}, b^\gamma_n \big)=b^\gamma_n$. In this case we have, $b^\gamma_n \max  \big( b_n \lambda_n^{-1} ,  b_n^{-1} \big) \geq b_n^{\gamma - 1}$ and the sequence on the right-hand side of this inequality converges to infinity for $\gamma <1$ and it is bounded for $\gamma =1$. Thus, in this case, it seems like we should take the value $\gamma = 1$ and the same sequences $(b_n)$, $(\lambda_n)$ as in Remark \ref{R21}. 
%(iii) Following the statements in (i) and (ii),  it seems that $\gamma =1$ is a good compromise. The %sequence    $(\lambda_n)_{n \in \N}$ can be that of  Remark \ref{R21} It will be necessary to study %by simulations what value should be taken for $\gamma$ so as not to lose the adaptability of the %penalty (the $\max$ in $ \widehat{\omega}_{n,j}$ between $n^{-1/2}$ and %$\sum^p_{k=1}|\widehat{\beta}_{j,k} - \widehat{\beta}_{j-1,k}|^\gamma$ is $n^{-1/2}$ for all $j \in %\{2, \cdots , g\}$) and that the estimation of ${\cal B}^0$ is as accurate as possible.
\end{remark}
 
Comparing now Theorem \ref{Proposition cardA} and Theorem \ref{Proposition cardA_bis}, we can deduce that the adaptive weights $\widehat{\omega}_{n,j}$ are responsible for a strong reduce the number of elements in $\overset{\vee}{{\cal B}_n} \cap \overline {{\cal B}^0}$, e.i, the false discoveries of different successive groups. This is also later confirmed by the simulation study performed in Section \ref{numerical}.

 \subsection{\textbf{\textit{Least squares loss function}}}
 \label{LS_sect}
In a standard linear regression model the least squares (LS) objective function is standardly used under the following assumptions imposed on the model errors:
 \begin{description}
\item  \textbf{(A4)} The error terms $(\varepsilon_i)_{1 \leqslant i \leqslant n}$ are i.i.d., such that $\E[\varepsilon]=0$ and $\Var[\varepsilon] < \infty$;
\end{description}

 We will now focus on the fused and adaptive fused group estimator based on the least squares objective function. In this case, instead of \eqref{Gn}, an analogous empirical process is considered
 \begin{equation}
 \label{Ln}
 L_n(\eb^g) \equiv \sum^n_{i=1}  (Y_i -\eeX_i^\top \eb^g)^2,
 \end{equation}
 with the corresponding estimator given as  
 $$\widetilde{{\eb}^{g}}_{\!\!(LS)}\equiv \argmin_{\eb^g \in \R^{r_n}}L_n(\eb^g),$$ 
and the penalized process analogous to \eqref{Qnb} is
 \begin{equation}
 \label{QUnb}
 U_n(\eb^g) \equiv L_n( \eb^g)+n \lambda_n \sum^g_{j=2} \| \eb_j-\eb_{j-1}\|_q,
 \end{equation}
with the corresponding  fused group LS estimator
$$\widehat{{\eb}^g}_{\!\!(LS)}\equiv  \argmin_{\eb^g \in \R^{r_n}} U_n(\eb^g).$$

A similar linear model with non-grouped explanatory variables ($p=1$) with the penalty of the form $\alpha \nu^{(1)}_n \sum^g_{j=1} | \eb_j|+(1-\alpha)\nu^{(1)}_n \sum_{j<k}| \eb_j-\eb_{k}|$, for some $\alpha \in (0,1]$,  with the LS objective function is also considered in \cite{Jang-Lim-Lazar-15}, however, for the situation where $g \in \mathbb{N}$ is fixed. The corresponding fused estimator allows the selection of groups of predictors that are positively correlated.

As already pointed out in Remark \ref{nouv_p1}, the results derived in this section for the LS framework are also novel for a model containing non-grouped variables ($p=1$) as the number of groups is allowed to increase with the sample size.

 The convergence rates of the proposed estimators  $\widetilde{{\eb}^{g}}_{\!\!(LS)}$  and $\widehat{{\eb}^g}_{\!\!(LS)}$ are the same as those obtained for the analogous estimators obtained for the quantile framework in Subsection \ref{Q_sect}.
 
 \begin{lemma}
 \label{Lemma 1_LS}
 Under Assumptions (A1), (A2), and (A4), and the sequence $(b_n)_{n \in \N}$ as in \eqref{abn},
 it holds that
$$\|\widetilde{{\eb}^{g}}_{\!\!(LS)} - \ebo\|_1=O_{\PP}(b_n). $$
 \end{lemma}

 Following the lines of the proof of Theorem \ref{theorem v_conv} we also obtain the proof of the following theorem.
 
 \begin{theorem}
 \label{theorem v_conv_LS}
Under Assumptions (A1), (A2), and (A4), the condition in \eqref{abn}, if $|{\cal B}^0|< \infty$ and $\lambda_nb_n^{-1} {\underset{n \rightarrow \infty}{\longrightarrow}} 0$, then  
$$\|  \widehat{{\eb}^g}_{\!\!(LS)} -\ebo\|_1=O_{\PP} (b_n).$$ 
\end{theorem}

The estimator of ${\cal B}^0$ based on $\widehat{{\eb}^g}_{\!\!(LS)}=\big( \widehat{\eb}_{1,(LS)}^\top, \cdots ,  \widehat{\eb}_{g,(LS)}^\top  \big)^\top$ is given in a straightforward way as
 \[
 \widehat{\cal B}_{n,(LS)}=\big\{ j \in \{2, \cdots , g \}; \widehat{\eb}_{j,(LS)} \neq \widehat{\eb}_{j-1,(LS)} \big\},
 \]
and a similar result to the one in Theorem \ref{Proposition cardA} can be derived again. 

 \begin{theorem}
 \label{Proposition cardA_LS}
 Under the same assumptions as in  Theorem \ref{theorem v_conv_LS},   there exists a positive constant $C_1$ such that, 
 \[ \lim_{n \rightarrow \infty} \PP\bigg[|    \widehat{\cal B}_{n,(LS)} \setminus {\cal B}^0| \leq C_1 \max  \bigg( \frac{b_n }{ \lambda_n} , \frac{1}{b_n} \bigg)\bigg]=1.
 \]
 \end{theorem}

Similarly as for the quantile framework before,  one can again improve the estimation accuracy of ${\cal B}^0$ by taking the advantage of $\widehat{{\eb}^g}_{\!\!(LS)}$ and defining the adaptive fused  penalty with the corresponding empirical process
 \begin{equation}
 \label{adfusedlasso}
 \overset{\vee}{ U}_n(\eb^g) \equiv L_n( \eb^g)+n \lambda_n \sum^g_{j=2} \widehat{\omega}_{n,j,(LS)} \| \eb_j-\eb_{j-1}\|_q,
 \end{equation}
 where the weights $\widehat{\omega}_{n,j,(LS)}$ are again constructed on the basis of fused group LS estimator as $\widehat{\omega}_{n,j,(LS)} \equiv 1/\max \big(n^{-1/2}, \sum^p_{k=1}|\widehat{\beta}_{j,k,(LS)} - \widehat{\beta}_{j-1,k,(LS)}|^\gamma \big)$, for  some fixed $\gamma>0$ and $\widehat{\beta}_{j,k,(LS)}$ being the $k$th component of $\widehat{\eb}_{j,(LS)}$. Thus, the adaptive fused group LS estimator is 
 \begin{equation*}
 \overset{\vee}{{\eb}^g}_{\!\!(aLS)}\equiv \argmin_{\eb^g \in \R^{r_n}} \overset{\vee}{ U}_n(\eb^g),
 \end{equation*} 
 and the corresponding estimator for ${\cal B}^0$ is defined as
  \[
  \overset{\vee}{{\cal B}}_{n,(aLS)} \equiv \big\{ j \in \{2, \cdots , g \}; \overset{\vee}{\eb}_{j,(aLS)} \neq \overset{\vee}{\eb}_{j-1,(aLS)} \big\}.
 \]
As for the quantile framework, the sequence $(\lambda_n)_{n \in \N}$, in relations (\ref{QUnb}) and (\ref{adfusedlasso}), can be different. 
Finally, using now the same arguments as in Theorem \ref{theorem v_conv_bis} and following the same lines of the proof, we obtain an analogous results also for $\overset{\vee}{{\eb}^g}_{\!\!(aLS)}$.

 \begin{theorem}
	\label{mm_theorem_added}
	Under Assumptions (A1), (A2), and (A4), the condition in \eqref{abn}, if $|{\cal B}^0|< \infty$, the for any sequence $(\lambda_n)_{n \in \N}$ such that $\lambda_nb_n^{-1} {\underset{n \rightarrow \infty}{\longrightarrow}} 0$, it holds that 
	$$\|\overset{\vee}{{\eb}^g}_{\!\!(aLS)} -\ebo\|_1=O_{\PP} (b_n).$$
	
\end{theorem}

 The results presented in Subsection \ref{Q_sect} and Subsection \ref{LS_sect} show that the estimated number of different successive groups is of the same order for both estimation frameworks with the  adaptive fused approach and, moreover, the convergence rates of the corresponding estimators for the model parameters are also of the same order, all under the assumption that the true number of groups is bounded. The finite sample performance is investigated in the next section.

\section{Numerical study and application}
\label{numerical}
In this section we firstly present a Monte Carlo simulation study to show some numerical properties of the proposed fused methods for the varying number of groups, different sample sizes and error distributions. Later, the application on the air quality data is presented. The goal is to detect daily moments when the temperature and humidity contribution change their effect with respect to the maximum daily benzene concentration.

\subsection{\textit{Numerical study}}

The fused group quantile estimator $\widehat{{\eb}^g}$ defined in terms of the minimization in \eqref{Qnb} and the adaptive  fused group quantile estimator $\overset{\vee}{{\eb}^g}$ in \eqref{quantile_adfuslasso} are both compared with the  fused group LS estimator $\widehat{{\eb}^g}_{\!\!(LS)}$ in \eqref{QUnb} and its adaptive version $\overset{\vee}{{\eb}^g}_{\!\!(aLS)}$  in \eqref{adfusedlasso} with respect to a wide range of different simulation settings. In order to make the comparison meaningful the quantile level of $\tau = 0.5$ is considered. The dimension of the unknown group specific vector of parameters $\boldsymbol{\beta} \in \mathbb{R}^p$ is either $p = 1$ or $p =3$ and three options are used for number of groups, $g \in \{20, 100, 200\}$. The sample is given as $n = p g$.  The model covariates are randomly generated from the normal distribution and two distributions are used for the error terms (standard normal and Cauchy). The true number of different successive groups  in the model is either $2$, $5$, $10$, or $n/5$, where the last option ($20~\%$ of the sample size) clearly does not satisfy the model assumptions but it is still included in the simulation setup for the comparison purposes. Obviously, if there are two change points in the group parameter then there are three successive groups. Analogously for $5$ changes in the group specific vector parameter---there are 6 successive groups.  The corresponding   locations of changes between successive groups are determined randomly and the jump magnitudes are also assigned randomly on the scale from $0.5$ to $2$ to allow for various signal-to-noise ratio. The regularization parameter equals $\lambda_n = n^{-1} (\log(n))^{1/2}$ 
for the fused method and $\lambda_n = n^{-1} (\log(n))^{5/2}$ for the adaptive fused method with the adaptive weights defined in \eqref{quantile_adfuslasso} and \eqref{adfusedlasso}  for $\gamma = 1$.

All four methods are compared with respect to the quality of the final fit and, mainly, the different successive coefficient group detection performance. The median  (MED) of $(Y_i - \widehat{Y}_i)_{1 \leqslant i \leqslant n}$ and the $L_1$  norm of the difference	
	 %%$(\boldsymbol{\beta}^0 - \widehat{\boldsymbol{\beta}})$ 
	 between the true vector of parameters and its estimate are used to evaluate the estimation performance while the true recovery rate (the proportion of truly detected different successive coefficient groups with respect to all unknown changes) and the overestimation rate (proportion of the number of detected different successive coefficient groups with respect to the number of true changes) are used to assess the detection performance. The results are summarized in Table \ref{tab1} (for $p = 1$) and Table \ref{tab2} (for $p = 3$).

For $M$ independent Monte Carlo replications let $\widehat{\eb}_{(m)}$ denote the estimate of $\eb^g$ by one of the four estimation methods for the $m$-th Monte Carlo run, with $m=1, \cdots , M$. The corresponding forecast for $Y_i$ is $\widehat{Y}_{i,(m)}=\eeX^\top_i\widehat{\eb}_{(m)} $. For each Monte Carlo replication  the median error $med_{(m)}=median (Y_i -\widehat{Y}_{i,(m)}; \; i=1,  \cdots , n)$ is obtained and the reported results are averaged over all $M$ simulation runs MED$ = M^{-1} \sum^M_{m=1} med_{(m)}$. For the parameter estimation the value of  MAD$ = M^{-1} \sum^M_{m=1} \frac{1}{p  g} \sum^{pg}_{j=1}| \beta^0_j -\widehat{\beta}_{j,(m)}|$ is reported.

\begin{figure}
	\centering
	\subfigure[Least Squares Error] {\label{fig0a}\includegraphics[width=0.48\textwidth, height = 4.5cm]{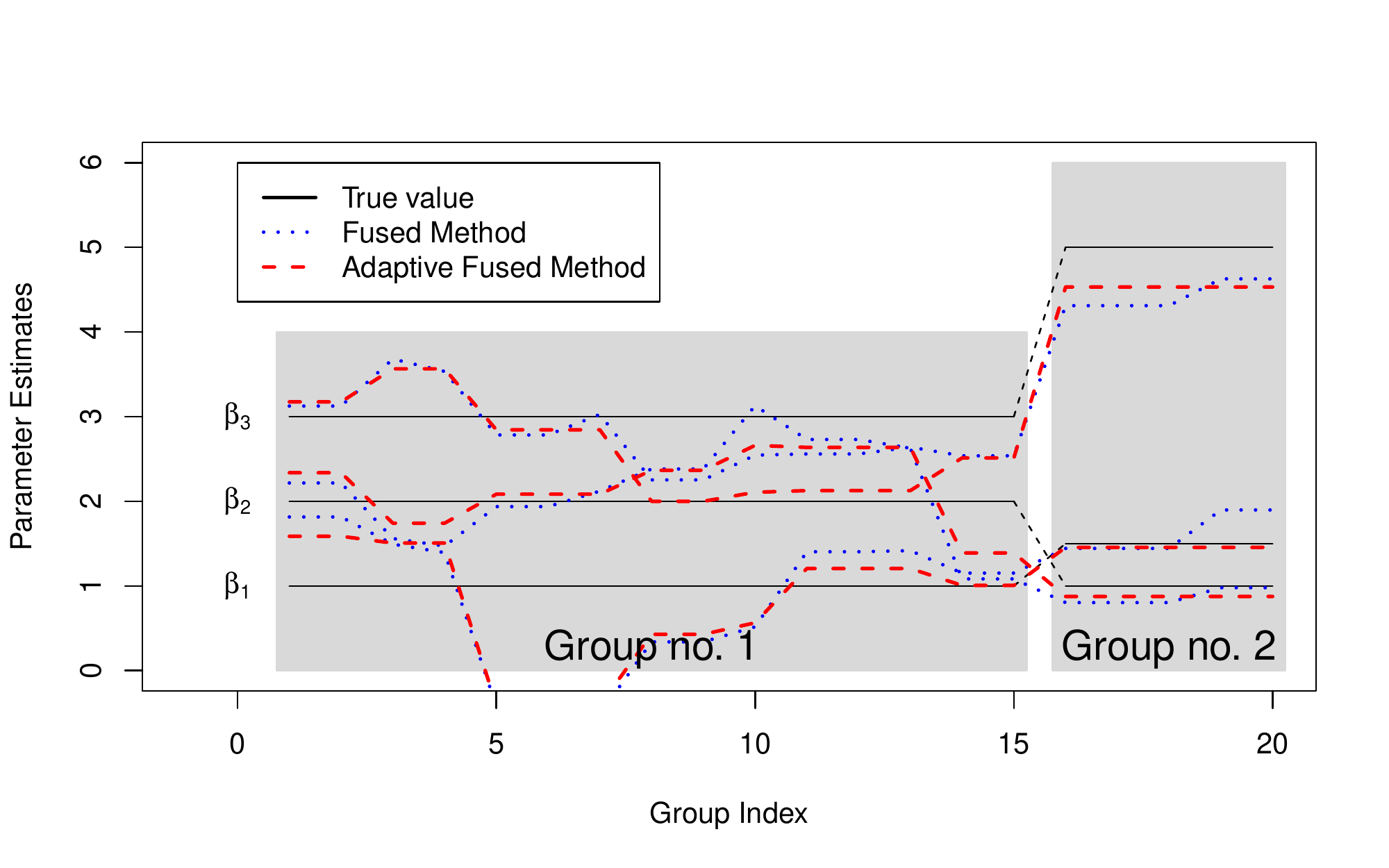}}
	\subfigure[Quantile Check Function] {\label{fig0b}\includegraphics[width=0.48\textwidth, height = 4.5cm]{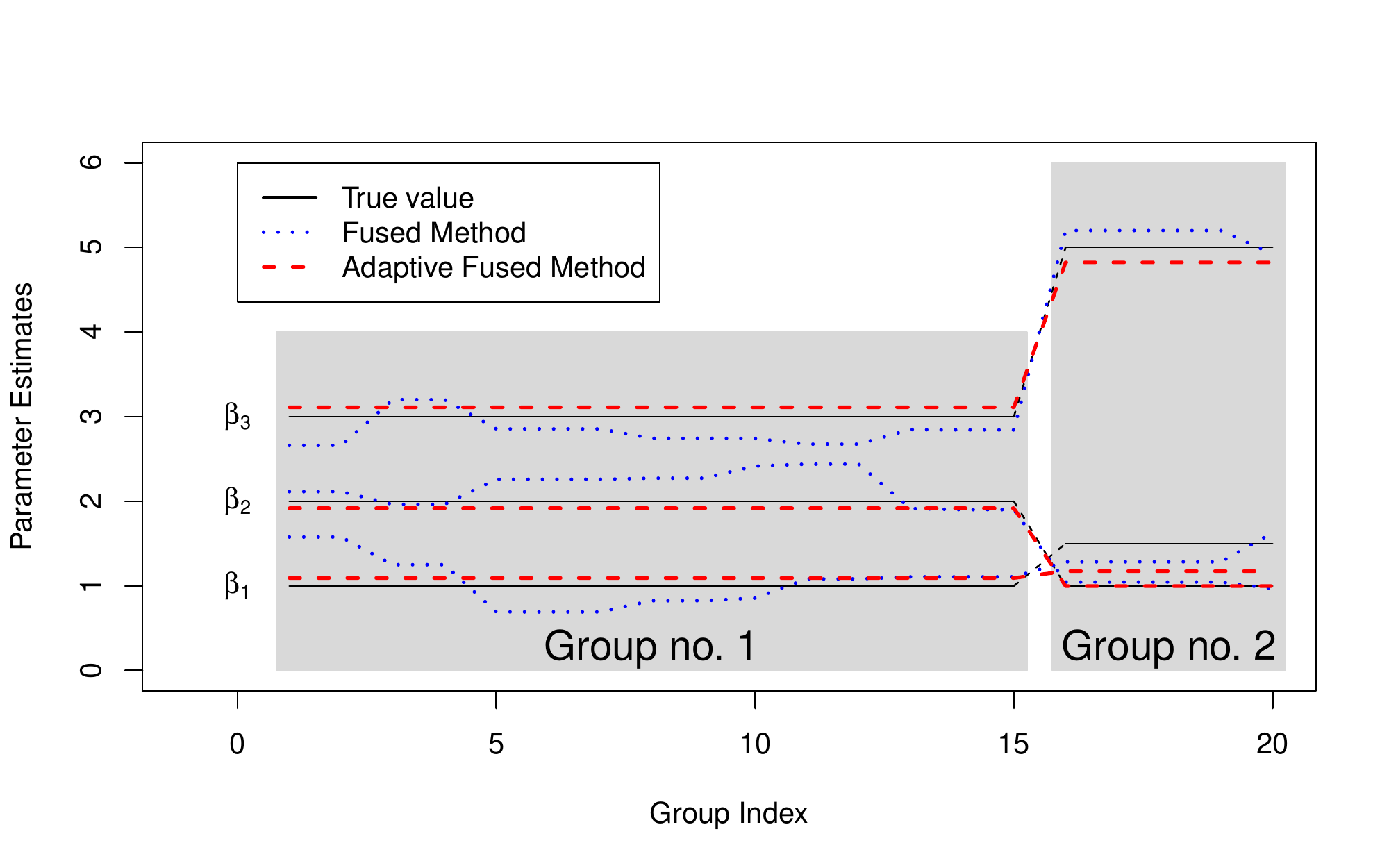}}
	
	\caption{\footnotesize An illustration of the model in \eqref{M1} for two truly different successive groups (out of $g = 20$ in total) and three explanatory variables in each group ($p = 3$). The first group specific vector parameter is the same for the groups $j \in \{1, \dots, 15\}$ and it differs from the second group specific vector parameter, which is the same for the groups $j \in \{16\dots, 20\}$. The Cauchy error terms are considered to visualize the robust favor of the quantile estimation approach for $\tau = 0.5$ (right panel) when compared with the standard least squares (left panel).} 
	\label{fig0}
\end{figure}

For some illustration of the model there is an example in Figure \ref{fig0}: the number of true different successive groups is two (out of $g = 20$ in total) and the number of the explanatory variables within each group is three ($p = 3$). The true vector of the group specific parameters for the first group (group indexes $j \in \{1, \dots, 15\}$) is $\boldsymbol{\beta}_{1}^{0} = (1,2,3)^\top$ and the true vector of the group specific parameters for the second group (group indexes $j \in \{16, \dots, 20\}$) is $\boldsymbol{\beta}_{1}^{0} = (1.5,1,5)^\top$. The sample size is $n = gp$ ($n = 80$). All four proposed  estimation methods are applied and the corresponding estimates are given in Figure \ref{fig0a} for \eqref{Ln} and \eqref{QUnb} and Figure \ref{fig0b} for \eqref{Gn} and \eqref{Qnb}.

\begin{landscape}
	\begin{table}[!ht]\footnotesize
		\begin{center}
			\scalebox{0.94}{\begin{tabular}{ccc|cccccccccccc}
\multirow{2}{*}{$\boldsymbol{\mathcal{D}}$} & & $\boldsymbol{p = 1}$ & 
     \multicolumn{3}{c}{\textbf{Model with 2 different groups}} & \multicolumn{3}{c}{\textbf{Model with 5 different groups}} & \multicolumn{3}{c}{\textbf{Model with 10 different groups}} & \multicolumn{3}{c}{\textbf{Model with $\boldsymbol{20~\%}$ different groups}} \\
 ~ & ~ & $\boldsymbol{g = n}$  & 
     \multicolumn{1}{c}{\scalebox{0.8}{MED}} & \multicolumn{1}{c}{\scalebox{0.8}{MAD}} & \multicolumn{1}{c}{\scalebox{0.8}{Recovery}} &
     \multicolumn{1}{c}{\scalebox{0.8}{MED}} & \multicolumn{1}{c}{\scalebox{0.8}{MAD}} & \multicolumn{1}{c}{\scalebox{0.8}{Recovery}} & 
     \multicolumn{1}{c}{\scalebox{0.8}{MED}} & \multicolumn{1}{c}{\scalebox{0.8}{MAD}} & \multicolumn{1}{c}{\scalebox{0.8}{Recovery}} &
     \multicolumn{1}{c}{\scalebox{0.8}{MED}} & \multicolumn{1}{c}{\scalebox{0.8}{MAD}} & \multicolumn{1}{c}{\scalebox{0.8}{Recovery}}  \\\hline\hline
\multicolumn{3}{c}{~} & \multicolumn{12}{c}{~}\\[-0.3cm]
$\boldsymbol{N}$ &  & \textbf{20} & 0.01 & 0.17 & 0.91$/$12.25&0.00 & 0.19 & 0.99$/$3.19&0.04 & 0.24 & 0.90$/$1.53&0.00 & 0.19 & 0.93$/$4.17 \\
 & \scalebox{0.8}{$\widehat{{\eb}^g}_{\!\!(LS)}$} & \textbf{100} & 0.00 & 0.09 & 1.00$/$71.48&-0.01 & 0.09 & 1.00$/$18.02&0.00 & 0.10 & 0.99$/$8.07&0.01 & 0.12 & 0.98$/$4.01 \\
 & & \textbf{200} & 0.01 & 0.07 & 1.00$/$152&0.00 & 0.07 & 1.00$/$38.24&0.00 & 0.07 & 1.00$/$17.14&0.00 & 0.08 & 1.00$/$4.07 \\
\multicolumn{3}{c}{~} & \multicolumn{12}{c}{~}\\[-0.2cm]
\rowcolor{Gray} \multicolumn{3}{c}{~} & \multicolumn{12}{c}{~}\\[-0.2cm]
\rowcolor{Gray} &  & \textbf{20} & -0.05 & 0.10 & 0.64$/$1.26&0.01 & 0.16 & 0.87$/$0.93&0.06 & 0.31 & 0.59$/$0.66&-0.04 & 0.14 & 0.72$/$0.83 \\
\rowcolor{Gray} & \scalebox{0.8}{$\overset{\vee}{{\eb}^g}_{\!\!(aLS)}$} & \textbf{100} & 0.00 & 0.02 & 0.93$/$1.84&-0.06 & 0.03 & 0.98$/$1.12&0.02 & 0.06 & 0.94$/$1.08&-0.12 & 0.12 & 0.81$/$0.94 \\
\rowcolor{Gray} &  & \textbf{200} & 0.00 & 0.01 & 0.99$/$2.15&0.02 & 0.01 & 1.00$/$1.13&-0.02 & 0.02 & 0.99$/$1.09&0.06 & 0.08 & 0.90$/$0.92 \\
\rowcolor{Gray} \multicolumn{3}{c}{~} & \multicolumn{12}{c}{~}\\[-0.2cm]
\rowcolor{white} \multicolumn{3}{c}{~} & \multicolumn{12}{c}{~}\\[-0.2cm]
~ &  & \textbf{20} & 0.01 & 0.12 & 0.88$/$6.44&-0.03 & 0.17 & 0.98$/$2.02&0.22 & 0.27 & 0.82$/$1.16&-0.01 & 0.15 & 0.91$/$2.59 \\
~ & \scalebox{0.8}{$\widehat{{\eb}^g}$} & \textbf{100} & 0.00 & 0.07 & 1.00$/$48.99&-0.02 & 0.07 & 1.00$/$12.67&0.01 & 0.08 & 0.99$/$5.86&0.02 & 0.10 & 0.97$/$3.06 \\
 & & \textbf{200} & 0.00 & 0.05 & 1.00$/$115&0.01 & 0.05 & 1.00$/$29.27&0.00 & 0.06 & 1.00$/$13.16&0.01 & 0.07 & 1.00$/$3.29 \\
\multicolumn{3}{c}{~} & \multicolumn{12}{c}{~}\\[-0.2cm]
\rowcolor{Gray} \multicolumn{3}{c}{~} & \multicolumn{12}{c}{~}\\[-0.2cm]
\rowcolor{Gray} &  & \textbf{20} & -0.22 & 0.18 & 0.00$/$0.00&-0.50 & 0.55 & 0.39$/$0.39&0.52 & 0.97 & 0.23$/$0.24&-0.12 & 0.55 & 0.31$/$0.31 \\
\rowcolor{Gray} &  \scalebox{0.8}{$\overset{\vee}{{\eb}^g}$}  & \textbf{100} & -0.05 & 0.08 & 0.82$/$0.91&-0.33 & 0.13 & 0.51$/$0.51&0.34 & 0.35 & 0.42$/$0.42&-0.34 & 0.47 & 0.18$/$0.18 \\
\rowcolor{Gray} & & \textbf{200} & 0.00 & 0.02 & 0.98$/$1.02&0.20 & 0.07 & 0.92$/$0.92&-0.15 & 0.20 & 0.84$/$0.84&0.24 & 0.45 & 0.44$/$0.44 \\
\rowcolor{Gray}\multicolumn{3}{c}{~} & \multicolumn{12}{c}{~}\\[-0.2cm]
\hline\rowcolor{white}\multicolumn{3}{c}{~} & \multicolumn{12}{c}{~}\\[-0.2cm]
$\boldsymbol{C}$ &  & \textbf{20} & 0.01 & 53.93 & 0.92$/$16.37&-0.12 & 60.79 & 0.93$/$4.05&0.09 & 19.27 & 0.89$/$1.82&0.03 & 25.92 & 0.92$/$5.42 \\
 & \scalebox{0.8}{$\widehat{{\eb}^g}_{\!\!(LS)}$} & \textbf{100} & 0.05 & 143 & 0.96$/$92.47&0.02 & 45.18 & 0.95$/$23.16&0.02 & 74.57 & 0.95$/$10.26&0.00 & 73.24 & 0.95$/$4.90 \\
 & & \textbf{200} & 0.00 & 38.55 & 0.97$/$190&0.01 & 34.36 & 0.96$/$47.66&-0.02 & 57.97 & 0.97$/$21.20&0.05 & 71.76 & 0.96$/$4.90 \\
\multicolumn{3}{c}{~} & \multicolumn{12}{c}{~}\\[-0.2cm]
\rowcolor{Gray} \multicolumn{3}{c}{~} & \multicolumn{12}{c}{~}\\[-0.2cm]
\rowcolor{Gray} &  & \textbf{20} & 1.72 & 62.95 & 0.63$/$10.50&1.54 & 81.97 & 0.73$/$2.74&1.82 & 20.97 & 0.67$/$1.31&1.70 & 27.58 & 0.71$/$3.60 \\
\rowcolor{Gray} & \scalebox{0.8}{$\overset{\vee}{{\eb}^g}_{\!\!(aLS)}$} & \textbf{100} & -1.51 & 456 & 0.74$/$68.01&-1.56 & 51.66 & 0.76$/$17.13&-1.48 & 88.17 & 0.76$/$7.64&-1.54 & 137 & 0.75$/$3.68 \\
\rowcolor{Gray} &  & \textbf{200} & -1.16 & 43.63 & 0.80$/$148&-1.14 & 36.66 & 0.79$/$37.09&-1.15 & 70.54 & 0.80$/$16.54&-1.13 & 88.10 & 0.80$/$3.88 \\
\rowcolor{Gray} \multicolumn{3}{c}{~} & \multicolumn{12}{c}{~}\\[-0.2cm]
\rowcolor{white} \multicolumn{3}{c}{~} & \multicolumn{12}{c}{~}\\[-0.2cm]
~ &  & \textbf{20} & 0.00 & 0.22 & 0.64$/$6.15&-0.05 & 0.35 & 0.84$/$1.90&0.33 & 0.49 & 0.66$/$0.97&-0.03 & 0.31 & 0.76$/$2.44 \\
~ & \scalebox{0.8}{$\widehat{{\eb}^g}$} & \textbf{100} & 0.00 & 0.13 & 0.85$/$48.71&-0.03 & 0.15 & 0.95$/$12.58&0.02 & 0.17 & 0.89$/$5.73&0.01 & 0.21 & 0.81$/$2.90 \\
 & & \textbf{200} & 0.00 & 0.12 & 0.93$/$115&0.00 & 0.13 & 0.98$/$29.12&-0.01 & 0.13 & 0.95$/$13.09&0.01 & 0.18 & 0.94$/$3.23 \\
\multicolumn{3}{c}{~} & \multicolumn{12}{c}{~}\\[-0.2cm]
\rowcolor{Gray} \multicolumn{3}{c}{~} & \multicolumn{12}{c}{~}\\[-0.2cm]
\rowcolor{Gray} &  & \textbf{20} & -0.22 & 0.20 & 0.01$/$0.01&-0.80 & 0.77 & 0.28$/$0.30&0.75 & 1.17 & 0.16$/$0.19&-0.41 & 0.74 & 0.20$/$0.21 \\
\rowcolor{Gray} &  \scalebox{0.8}{$\overset{\vee}{{\eb}^g}$}  & \textbf{100} & -0.04 & 0.11 & 0.41$/$1.03&-0.34 & 0.14 & 0.51$/$0.53&0.32 & 0.35 & 0.39$/$0.44&-0.34 & 0.47 & 0.19$/$0.20 \\
\rowcolor{Gray} & & \textbf{200} & -0.04 & 0.04 & 0.62$/$1.96&0.30 & 0.11 & 0.65$/$0.69&-0.16 & 0.26 & 0.60$/$0.74&0.38 & 0.52 & 0.39$/$0.40 \\
\rowcolor{Gray}\multicolumn{3}{c}{~} & \multicolumn{12}{c}{~}\\[-0.2cm]\hline\hline

     \rowcolor{white}\multicolumn{3}{c}{~} & \multicolumn{12}{c}{~}\\[-0.4cm]\hline

     \end{tabular}}	
		\end{center}
		\caption{\footnotesize Simulation results for the situation where the dimension of the unknown parameter is $p = 1$. Two goodness-of-fit quantities are provided: the median (MED) of $(Y_i - \widehat{Y}_i)_{1 \leqslant i \leqslant n}$ and the mean absolute difference (MAD) between the true parameter vector $\boldsymbol{\beta}^0$ and the corresponding empirical estimate $\widehat{\boldsymbol{\beta}}$.  The \textit{Recovery} column is given in terms of two values: the proportion of truly discovered different successive coefficients (value $1$ stands for all true changes being discovered) and the proportion between the number of estimated different successive coefficients and true changes (value $1$ stands for the situation where the number of estimated changes equals the number of true changes). An ideal situation is $1.00/1.00$ which means that all true changes are discovered with no other detections in addition. The results are averaged over 1000 Monte Carlo simulation runs.}
		\label{tab1}
	\end{table}   
\end{landscape}

\begin{landscape}
	\begin{table}[!ht]\footnotesize
		\begin{center}
			\scalebox{0.94}{\begin{tabular}{ccc|cccccccccccc}
\multirow{2}{*}{$\boldsymbol{\mathcal{D}}$} & & $\boldsymbol{p = 3}$ & 
     \multicolumn{3}{c}{\textbf{Model with 2 different groups}} & \multicolumn{3}{c}{\textbf{Model with 5 different groups}} & \multicolumn{3}{c}{\textbf{Model with 10 different groups}} & \multicolumn{3}{c}{\textbf{Model with $\boldsymbol{20~\%}$ different groups}} \\
 ~ & ~ & $\boldsymbol{g = n/p}$  & 
     \multicolumn{1}{c}{\scalebox{0.8}{MED}} & \multicolumn{1}{c}{\scalebox{0.8}{MAD}} & \multicolumn{1}{c}{\scalebox{0.8}{Recovery}} &
     \multicolumn{1}{c}{\scalebox{0.8}{MED}} & \multicolumn{1}{c}{\scalebox{0.8}{MAD}} & \multicolumn{1}{c}{\scalebox{0.8}{Recovery}} & 
     \multicolumn{1}{c}{\scalebox{0.8}{MED}} & \multicolumn{1}{c}{\scalebox{0.8}{MAD}} & \multicolumn{1}{c}{\scalebox{0.8}{Recovery}} &
     \multicolumn{1}{c}{\scalebox{0.8}{MED}} & \multicolumn{1}{c}{\scalebox{0.8}{MAD}} & \multicolumn{1}{c}{\scalebox{0.8}{Recovery}}  \\\hline\hline
\multicolumn{3}{c}{~} & \multicolumn{12}{c}{~}\\[-0.3cm]
$\boldsymbol{N}$ & & \textbf{20} & -0.03 & 0.17 & 1.00$/$9.87&0.02 & 0.40 & 0.92$/$2.63&0.02 & 1.00 & 0.65$/$1.11&-0.01 & 0.34 & 0.76$/$3.37 \\
 & \scalebox{0.8}{$\widehat{{\eb}^g}_{\!\!(LS)}$} & \textbf{100} & 0.00 & 0.05 & 1.00$/$62.67&0.00 & 0.06 & 1.00$/$15.61&0.00 & 0.08 & 1.00$/$6.72&0.00 & 0.14 & 1.00$/$3.03 \\
 & & \textbf{200} & 0.00 & 0.02 & 1.00$/$87.79&0.00 & 0.03 & 1.00$/$20.40&-0.07 & 0.07 & 1.00$/$11.03&0.00 & 0.10 & 1.00$/$2.50 \\
\multicolumn{3}{c}{~} & \multicolumn{12}{c}{~}\\[-0.2cm]
\rowcolor{Gray} \multicolumn{3}{c}{~} & \multicolumn{12}{c}{~}\\[-0.2cm]
\rowcolor{Gray} &  & \textbf{20} & -0.02 & 0.08 & 0.99$/$1.82&0.07 & 0.44 & 0.63$/$1.31&-0.02 & 1.09 & 0.39$/$0.64&-0.05 & 0.37 & 0.41$/$1.65 \\
\rowcolor{Gray} & \scalebox{0.8}{$\overset{\vee}{{\eb}^g}_{\!\!(aLS)}$} & \textbf{100} & 0.00 & 0.01 & 1.00$/$2.00&0.00 & 0.02 & 1.00$/$1.13&0.00 & 0.03 & 1.00$/$1.04&0.04 & 0.08 & 0.97$/$1.12 \\
\rowcolor{Gray} &  & \textbf{200} & 0.00 & 0.01 & 1.00$/$1.26&0.00 & 0.01 & 1.00$/$1.02&0.00 & 0.02 & 1.00$/$1.17&0.02 & 0.05 & 0.99$/$1.02 \\
\rowcolor{Gray} \multicolumn{3}{c}{~} & \multicolumn{12}{c}{~}\\[-0.2cm]
\rowcolor{white} \multicolumn{3}{c}{~} & \multicolumn{12}{c}{~}\\[-0.2cm]
~ &  & \textbf{20} & -0.04 & 0.18 & 1.00$/$9.92&0.00 & 0.41 & 0.92$/$2.70&0.08 & 1.01 & 0.57$/$1.02&-0.01 & 0.35 & 0.77$/$3.48 \\
~ & \scalebox{0.8}{$\widehat{{\eb}^g}$} & \textbf{100} & -5.79 & 2.05 & 1.00$/$99.00&-8.29 & 2.47 & 1.00$/$24.75&-4.86 & 0.91 & 1.00$/$10.08&0.05 & 0.17 & 1.00$/$4.28 \\
 & & \textbf{200} & -0.01 & 0.02 & 1.00$/$29.92&0.00 & 0.02 & 1.00$/$8.44&-0.02 & 0.03 & 1.00$/$4.73&-0.06 & 0.36 & 0.82$/$1.76 \\
\multicolumn{3}{c}{~} & \multicolumn{12}{c}{~}\\[-0.2cm]
\rowcolor{Gray} \multicolumn{3}{c}{~} & \multicolumn{12}{c}{~}\\[-0.2cm]
\rowcolor{Gray} &  & \textbf{20} & 0.03 & 0.24 & 0.85$/$0.90&0.00 & 0.53 & 0.33$/$0.63&-0.16 & 1.15 & 0.29$/$0.38&-0.07 & 0.61 & 0.22$/$0.89 \\
\rowcolor{Gray} & \scalebox{0.8}{$\overset{\vee}{{\eb}^g}$} & \textbf{100} & 0.06 & 0.07 & 0.07$/$2.57&-0.09 & 0.13 & 0.80$/$3.06&-0.21 & 0.15 & 0.88$/$1.31&-0.07 & 0.38 & 0.70$/$0.72 \\
\rowcolor{Gray} &  & \textbf{200} & 0.00 & 0.01 & 1.00$/$1.00&-0.01 & 0.01 & 1.00$/$1.00&0.00 & 0.02 & 1.00$/$1.00&-0.15 & 0.51 & 0.51$/$0.69 \\
\rowcolor{Gray}\multicolumn{3}{c}{~} & \multicolumn{12}{c}{~}\\[-0.2cm]
\hline\rowcolor{white}\multicolumn{3}{c}{~} & \multicolumn{12}{c}{~}\\[-0.2cm]
$\boldsymbol{C}$ & & \textbf{20} & 0.00 & 3.24 & 0.73$/$10.85&0.01 & 3.26 & 0.70$/$2.73&0.01 & 3.54 & 0.65$/$1.21&0.00 & 3.05 & 0.65$/$3.67 \\
 & \scalebox{0.8}{$\widehat{{\eb}^g}_{\!\!(LS)}$} & \textbf{100} & -0.08 & 3.36 & 0.86$/$71.60&0.05 & 3.18 & 0.83$/$17.35&-0.22 & 3.12 & 0.82$/$7.64&-0.01 & 3.16 & 0.81$/$3.55 \\
 & & \textbf{200} & -0.03 & 2.29 & 0.79$/$128&-0.10 & 2.29 & 0.82$/$31.83&-0.12 & 2.30 & 0.80$/$14.52&-0.16 & 2.42 & 0.79$/$3.24 \\
\multicolumn{3}{c}{~} & \multicolumn{12}{c}{~}\\[-0.2cm]
\rowcolor{Gray} \multicolumn{3}{c}{~} & \multicolumn{12}{c}{~}\\[-0.2cm]
\rowcolor{Gray} &  & \textbf{20} & 1.69 & 3.54 & 0.57$/$6.26&1.76 & 3.45 & 0.45$/$1.64&1.74 & 3.91 & 0.45$/$0.78&1.71 & 3.37 & 0.42$/$2.20 \\
\rowcolor{Gray} & \scalebox{0.8}{$\overset{\vee}{{\eb}^g}_{\!\!(aLS)}$} & \textbf{100} & -2.04 & 3.60 & 0.68$/$35.92&-1.65 & 3.32 & 0.62$/$9.17&-1.58 & 3.29 & 0.62$/$4.23&-1.53 & 3.42 & 0.58$/$1.99 \\
\rowcolor{Gray} &  & \textbf{200} & -1.22 & 2.49 & 0.67$/$75.66&-1.11 & 2.49 & 0.67$/$18.97&-1.17 & 2.54 & 0.65$/$8.49&-1.16 & 2.68 & 0.62$/$2.01 \\
\rowcolor{Gray} \multicolumn{3}{c}{~} & \multicolumn{12}{c}{~}\\[-0.2cm]
\rowcolor{white} \multicolumn{3}{c}{~} & \multicolumn{12}{c}{~}\\[-0.2cm]
~ &  & \textbf{20} & -0.02 & 0.55 & 0.76$/$9.84&0.02 & 0.71 & 0.65$/$2.53&0.05 & 1.17 & 0.56$/$1.09&0.00 & 0.73 & 0.62$/$3.40 \\
~ & \scalebox{0.8}{$\widehat{{\eb}^g}$} & \textbf{100} & -8.20 & 2.03 & 1.00$/$97.88&-9.66 & 2.30 & 1.00$/$24.61&-7.85 & 1.62 & 0.98$/$10.38&-0.63 & 0.67 & 0.97$/$4.89 \\
 & & \textbf{200} & -2.66 & 0.39 & 1.00$/$109&-2.45 & 0.40 & 1.00$/$27.59&-1.27 & 0.22 & 1.00$/$6.18&-1.12 & 0.57 & 0.77$/$1.87 \\
\multicolumn{3}{c}{~} & \multicolumn{12}{c}{~}\\[-0.2cm]
\rowcolor{Gray} \multicolumn{3}{c}{~} & \multicolumn{12}{c}{~}\\[-0.2cm]
\rowcolor{Gray} &  & \textbf{20} & -0.02 & 0.48 & 0.32$/$1.34&0.12 & 0.71 & 0.22$/$0.71&-0.17 & 1.28 & 0.22$/$0.35&0.02 & 0.75 & 0.23$/$1.00 \\
\rowcolor{Gray} & \scalebox{0.8}{$\overset{\vee}{{\eb}^g}$} & \textbf{100} & -0.11 & 0.12 & 0.48$/$9.37&-0.56 & 0.26 & 0.67$/$5.14&-0.58 & 0.39 & 0.49$/$1.61&-0.24 & 0.60 & 0.37$/$0.71 \\
\rowcolor{Gray} &  & \textbf{200} & -0.52 & 0.06 & 0.99$/$8.28&-0.26 & 0.07 & 0.89$/$2.41&-0.41 & 0.11 & 0.95$/$1.48&-0.32 & 0.60 & 0.43$/$0.66 \\
\rowcolor{Gray}\multicolumn{3}{c}{~} & \multicolumn{12}{c}{~}\\[-0.2cm]\hline\hline

     \rowcolor{white}\multicolumn{3}{c}{~} & \multicolumn{12}{c}{~}\\[-0.4cm]\hline

     \end{tabular}}	
		\end{center}
		\caption{\footnotesize Simulation results for the situation where the dimension of the unknown parameter is $p = 3$.  Two goodness-of-fit quantities are provided: the median (MED) of $(Y_i - \widehat{Y}_i)_{1 \leqslant i \leqslant n}$ and the mean absolute difference (MAD) between the true parameter vector $\boldsymbol{\beta}^0$ and the corresponding empirical estimate $\widehat{\boldsymbol{\beta}}$.  The \textit{Recovery} column is given in terms of two values: the proportion of truly discovered different successive coefficient groups (value $1$ stands for all true changes being discovered) and the proportion between the number of estimated different successive coefficient groups and true changes (value $1$ stands for the situation where the number of estimated changes equals the number of true changes). An ideal situation is $1.00/1.00$ which means that all true changes are discovered with no other detections in addition. The results are averaged over 1000 Monte Carlo simulation runs.}
		\label{tab2}
	\end{table}   
\end{landscape}

From Tables \ref{tab1} and \ref{tab2} for the Gaussian errors, we deduce that for $| {\cal B}^0| \in\{2, 5, 10 \}$, the fused estimations for the least squares and the quantile methods have the same properties and the same also applies for the adaptive frameworks which, moreover, have  the recovery detection rates close to one. On the other hand, if the assumption $|{\cal B}^0|< \infty$ does not hold, as for the models with $20\%$ different successive groups, then not all of the different successive groups are detected and the performance is worse. 

The robustness of the quantile methods is obvious when the Cauchy errors are used instead: while the LS based methods fail in both, the estimation and the group detection, the quantile approaches perform comparably well as in the situations with the Gaussian errors.

 \subsection{\textit{Application to Air Quality data}}
 \label{real_data}

In order to demonstrate the practical applicability of the proposed  model we use the air quality data from \cite{devito} which can be downloaded from the \textit{Machine Learning Repository} site \href{http://archive.ics.uci.edu/ml/datasets/Air+Quality\#}{http://archive.ics.uci.edu/ml/datasets/Air+Quality\#}.  The hourly meteorological and air quality data were recorded from March 2004 to February 2005.
The idea is to use the daily temperature and humidity profiles (recorded every hours) to predict the maximum benzene concentration level for the given day.  Optimally, it would be appropriate to use the temperature and humidity information only from some few instant moments during the day instead of recording both continuous profiles over the whole day. Given the data, there are $g = 24$ hourly groups and for each group there is the corresponding vector parameter $\boldsymbol{\beta}_j = (\beta_j^{T}, \beta_j^{H})^\top \in \mathbb{R}^2$, for $j = 1, \dots, 24$, where $\beta_{j}^{T}$ is responsible for the contribution of the temperature at '$j$' o'clock and $\beta_{j}^{H}$ models the effect of the humidity, again at '$j$' o'clock. Using the model formulation from Section \ref{Section_Model} and the estimation in terms of \eqref{Qnb} it can be achieved that most of the corresponding parameter vector estimates  are the same. If otherwise, then the existing changes in the vector estimates identify some specific daily segments with the same temperature and humidity contribution with respect to the maximum daily benzene concentration. The corresponding magnitudes for both effects in each daily segment are all estimated simultaneously.

Similarly as in the simulation section, four different models are fitted: the fused group LS approach and its adaptive version both presented in Figures \ref{fig2a} and \ref{fig2a} and the proposed fused group quantile and the adaptive fused group quantile in Figures \ref{fig2c} and \ref{fig2d}. The temperature data and the humidity data are heavily skewed and, therefore, it can be assumed that robust approaches are more appropriate for this situation. 

\begin{figure}
	\centering
	\subfigure[Daily temperature] {\label{fig1a}\includegraphics[width=0.48\textwidth, height = 4.5cm]{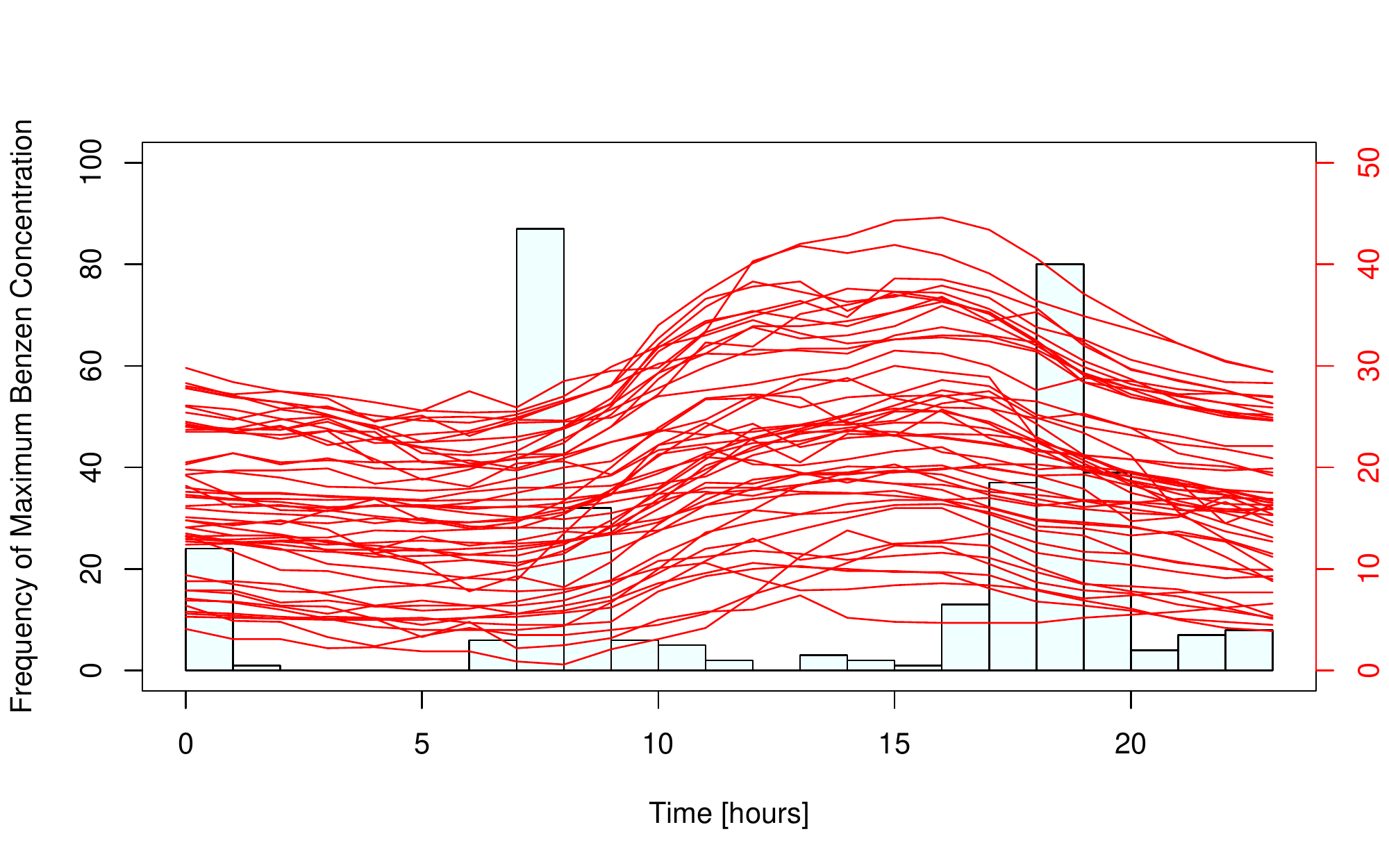}}
	\subfigure[Daily humidity] {\label{fig1b}\includegraphics[width=0.48\textwidth, height = 4.5cm]{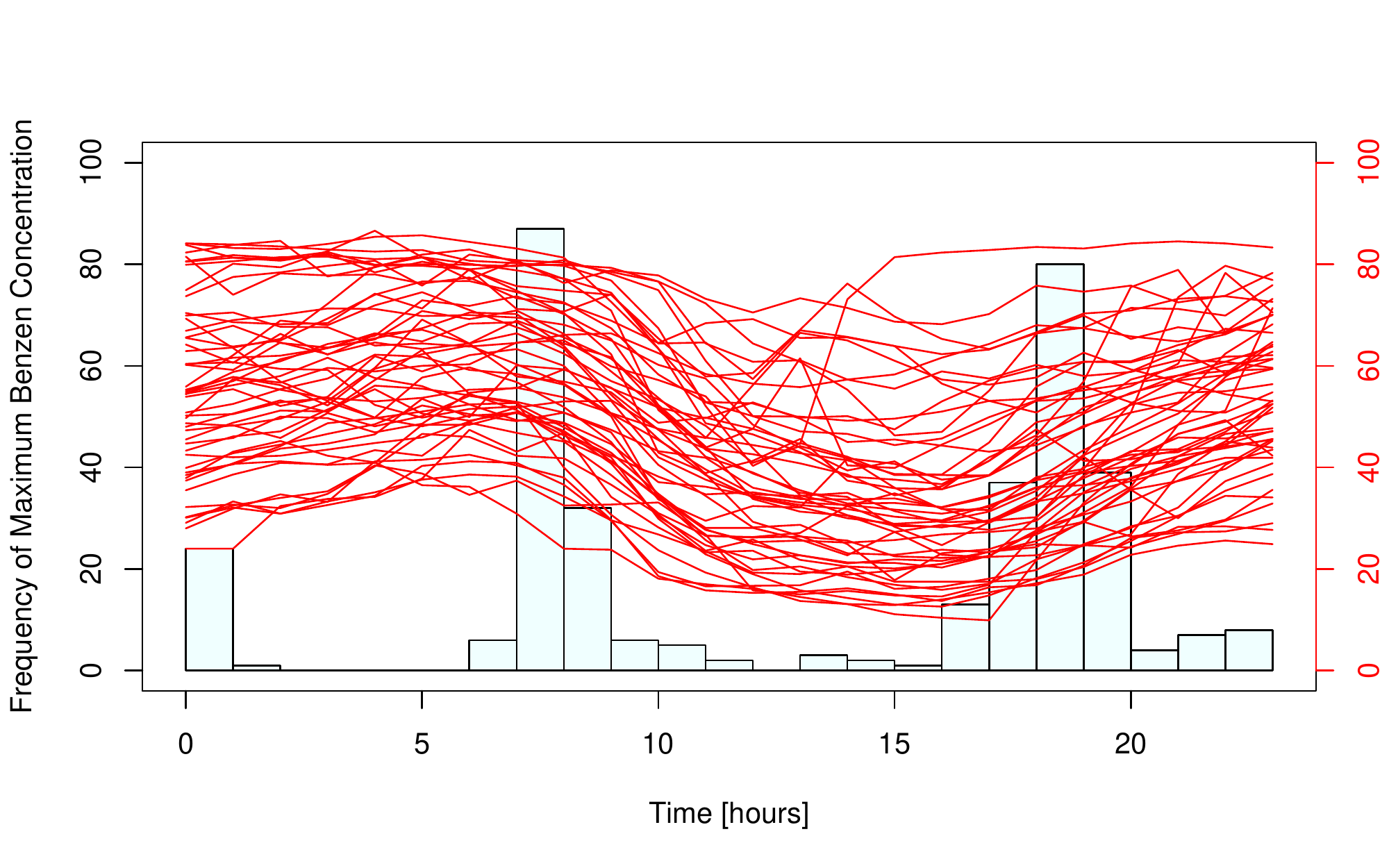}}
	
	\caption{\footnotesize Daily temperature profiles (left panel) and daily humidity profiles (right panel) for 50 randomly selected days out of 357 available days with full profiles in total. In addition, the maximum benzene concentration is recorded for each day and the corresponding time of the maximum occurrence (in hours) is given in terms of the frequency histograms in both panels.}
	\label{fig1}
\end{figure}

\begin{figure}
	\centering
	\subfigure[Fused group LS solution] {\label{fig2a}\includegraphics[width=0.48\textwidth, height = 4.5cm]{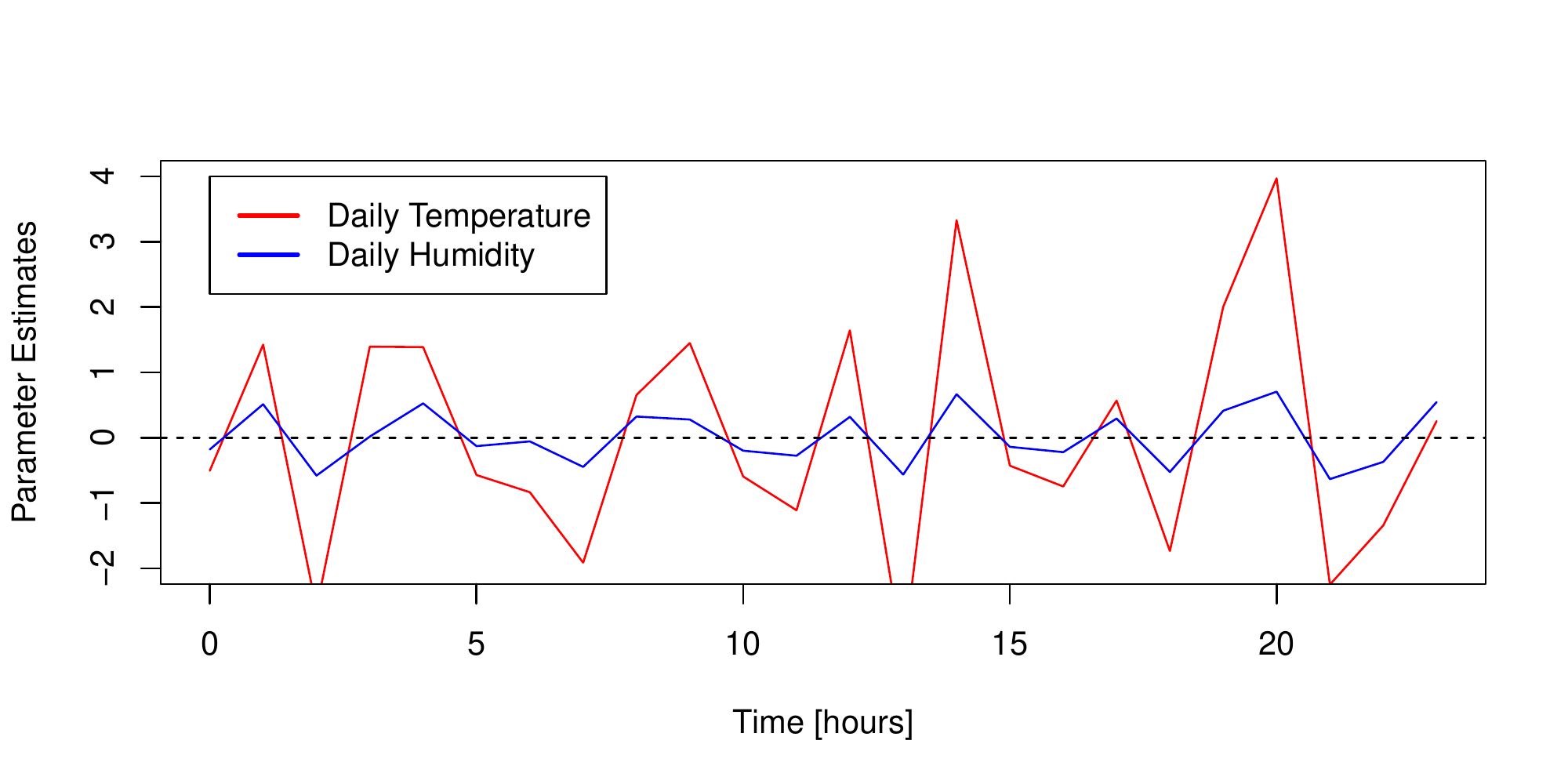}}
	\subfigure[Adaptive fused group LS solution] {\label{fig2b}\includegraphics[width=0.48\textwidth, height = 4.5cm]{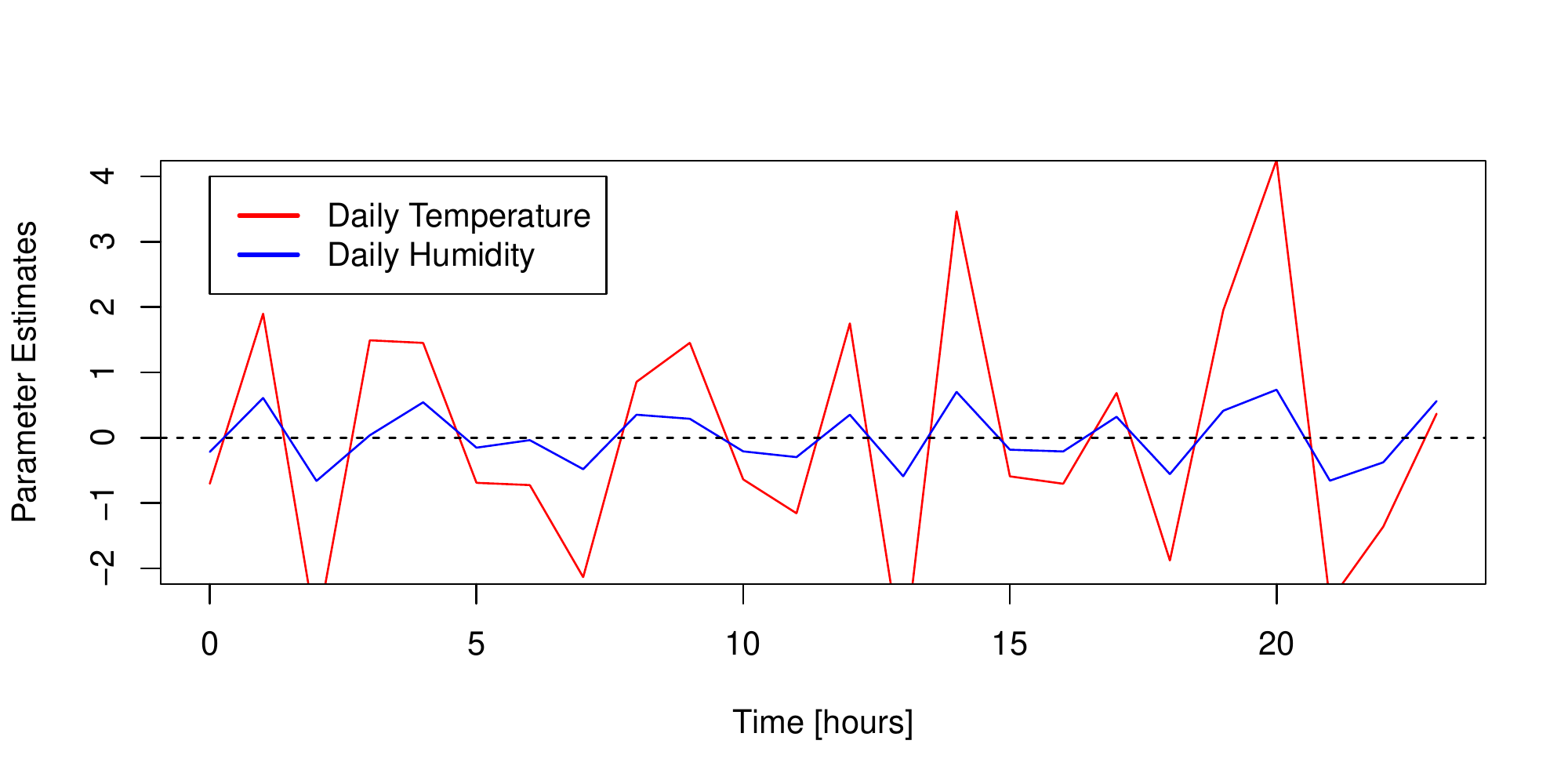}}
	
	\subfigure[Fused group quantile solution] {\label{fig2c}\includegraphics[width=0.48\textwidth, height = 4.5cm]{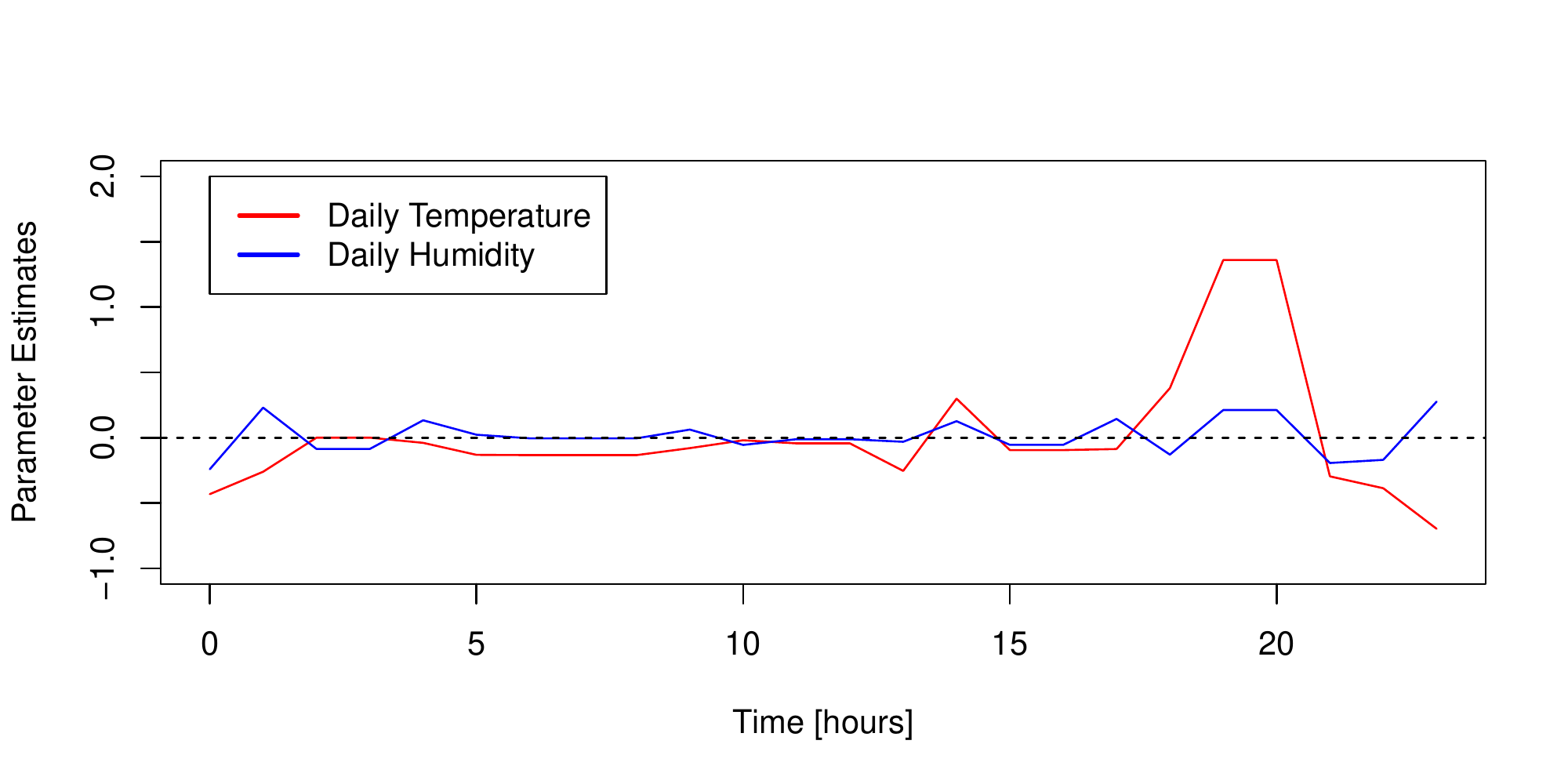}}
	\subfigure[Adaptive fused group quantile solution] {\label{fig2d}\includegraphics[width=0.48\textwidth, height = 4.5cm]{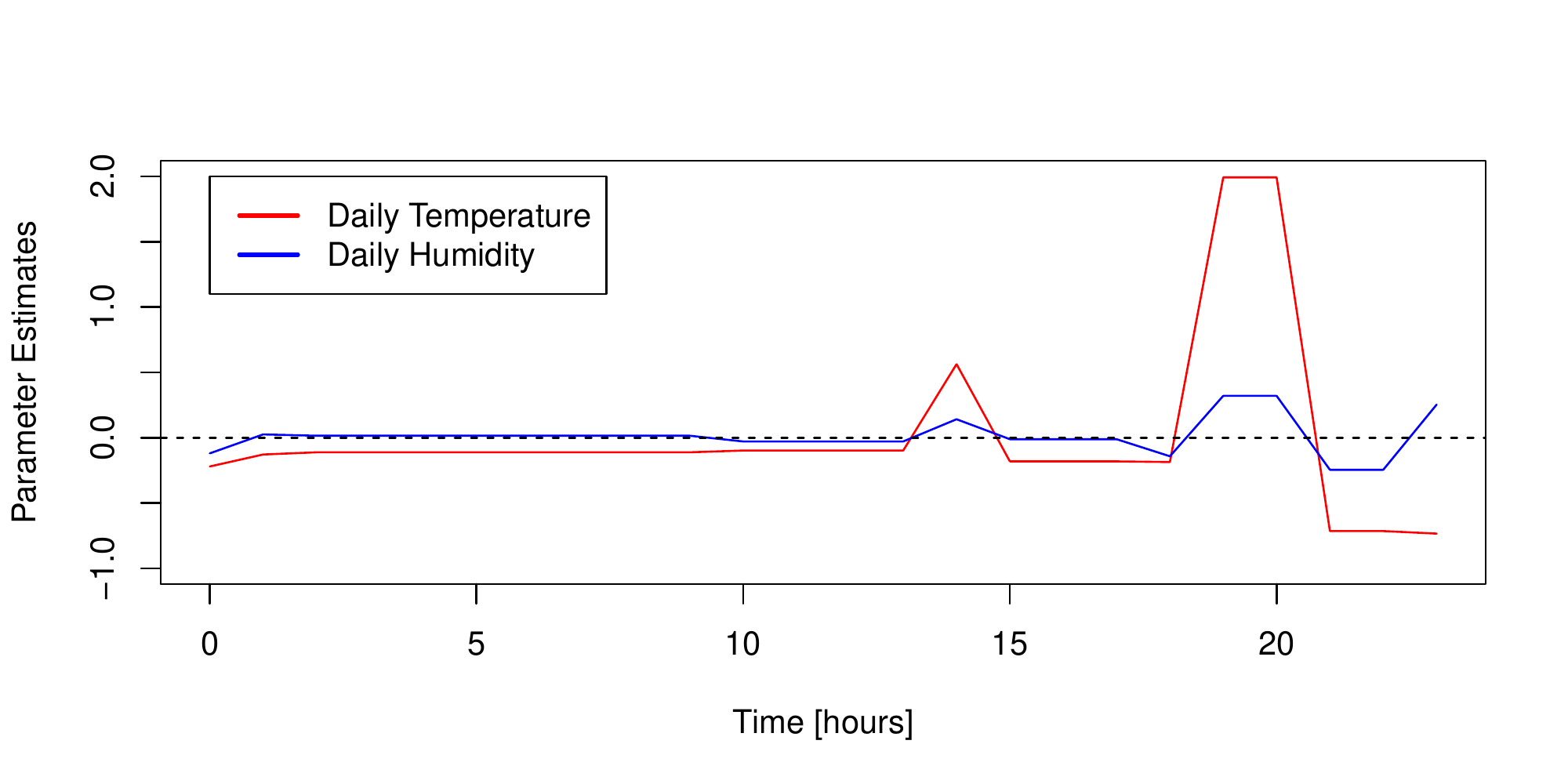}}

	\caption{\footnotesize The estimated parameter vectors $\widehat{\boldsymbol{\beta}}_j = (\widehat{\beta}_j^{T}, \widehat{\beta}_j^{H})^\top \in \mathbb{R}^2$, for $j = 1, \dots, 24$, for four different estimation techniques: fused group LS, adaptive fused group LS, fused group quantile and adaptive fused group quantile. The adaptive fused group quantile estimator in panel (d) clearly identifies some instant moments during a day when the temperature and humidity information is relevant for the maximum benzene concentration. In other words, it seems enough to record the temperature and humidity information at 2~pm and, also, after 6~pm. }
	\label{fig2}
\end{figure}

Indeed, while the fused group LS and also its adaptive version can not identify any specific daily moments which should be used to determine the maximum daily benzene concentration, the fused group quantile and its adaptive version in particular clearly identify some segments during the day when the contribution of the temperature and humidity is obvious.

  \section{Proofs}
  \label{Section_Proofs}
Throughout the proofs,  the following  identity  for the quantile check function $\rho_\tau$ is be used: for any $x,y \in \R$ it holds that
\begin{equation}
\label{rho}
\rho_\tau(x-y)- \rho_\tau (x)=y(\e1_{x<0} - \tau )+ \int^{\tau}_0 (\e1_{x \leq v} - \eu_{x \leq 0})dv.
\end{equation}
 \noindent {\bf Proof of Lemma \ref{Lemma 1}}.\\
 We will show that for all $\epsilon >0$, there exists a constant  $  C_\epsilon >0$, such that for $n$ large enough, we have
 \begin{equation}
 \label{eq20}
 \PP \left[ \inf_{\eu \in \R^{r_n}, \| \eu\|_1=1 } G_n \big( \ebo+ C_\epsilon b_n \eu \big) > G_n(\ebo)   \right] \geq 1 - \epsilon.
 \end{equation}
  Then, for any constant $c_1>0$, we can write the difference $G_n \big( \ebo+ c_1 b_n \eu \big) - G_n(\ebo) $ using the form
 \begin{equation}
 \label{GWR}
 G_n \big( \ebo+ c_1 b_n \eu \big) - G_n(\ebo) = \E \big[G_n \big( \ebo+c_1 b_n \eu \big) - G_n(\ebo)   \big]+\bfW_n^\top \eu +\sum^n_{i=1} ({\cal R}_i -\E[{\cal R}_i]) ,
 \end{equation}
 with  the   $r_n$-dimensional random vector $\bfW_n \equiv c_1 b_n \sum^n_{i=1} {\cal D}_i \eeX_i$, the random variables  ${\cal D}_i  \equiv  (1-\tau) \e1_{\{\varepsilon_i <0\}}- \tau \e1_{\{\varepsilon_i \geq 0\}}$, and   ${\cal R}_i \equiv \rho_\tau(\varepsilon_i - c_1 b_n \eeX_i^\top \eu)- \rho_\tau (\varepsilon_i) - c_1 b_n {\cal D}_i \eeX_i^\top \eu $.
 
Using the  Holder's inequality, we have that $|\eeX_i ' \eu | \leq \| \eeX_i\|_{\infty}  \| \eu \| _1$. Then, for all $\eu \in \R^{r_n}$ such that $\|\eu\|_1 = 1$, by Assumption (A1), we have that  $|\eeX_i^\top \eu| \leq C$. 

Firstly, we study the first term on the right-hand side  of relation \eqref{GWR}. Using the identity in  \eqref{rho}, we obtain
\[
G_n \big( \ebo+c_1 b_n \eu \big) - G_n(\ebo)= - c_1 b_n \sum^n_{i=1}\eeX_i^\top \eu {\cal D}_i+\sum^n_{i=1} \int_0^{c_1 b_n \eeX_i^\top \eu} [\e1_{\{\varepsilon_i < v\}} -\e1_{\{\varepsilon_i <0\}} ]dv .
\]
Applying now the mean value theorem, taking into account the fact that the derivative of $f$  is bounded in a neighborhood of zero by  Assumption (A3), and the fact that $\|\eu\|_1=1$, $\E[{\cal D}_i]=0$ and $b_n \rightarrow 0$, we obtain
 \begin{align*}
 \E \big[G_n \big( \ebo+c_1 b_n \eu \big) - G_n(\ebo)   \big] &=\sum^n_{i=1}\E\Big[\int_0^{c_1 b_n \eeX_i^\top \eu} \e1_{\{0 < \varepsilon_i < v\}} \Big] dv\\
 & = \sum^n_{i=1} \int_0^{c_1 b_n \eeX_i^\top \eu} [F(v)-F(0)]dv \\
& =\frac{f(0)}{2} c^2_1 b^2_n \sum^n_{i=1} (\eeX_i^\top \eu)^2 +o \Big(b^2_n \sum^n_{i=1} \eu^\top (\eeX_i \eeX_i^\top) \eu \Big).
 \end{align*}
 Using Assumption (A2) together with {$f(0)>0$, we get that
 \begin{equation}
 \label{eg}
  n^{-1}\E \big[G_n \big( \ebo+ c_1 b_n \eu \big) - G_n(\ebo)   \big] = C f(0) b^2_n \frac{1}{n} \sum^n_{i=1} \eu^\top \eeX_i \eeX^\top_i \eu (1+o(1)) >0.
 \end{equation}
 Next, we study the last two terms on the right-hand side  of relation \eqref{GWR}.
For the last term we have, with probability  one, for any $i=1, \dots , n$, that $|{\cal R}_i  | \leq c_1 b_n |\eeX_i^\top \eu| \e1_{\{|\varepsilon_i| \leq c_1 b_n |\eeX_i^\top \eu|\}}$. Since $(\varepsilon_i)_{1\leqslant i \leqslant n}$ are independent, then the random variables $({\cal R}_i)_{1\leqslant i \leqslant n}$ are independent as  well and, therefore 
 \begin{align}
 \label{eq140}
 \E \big[ \sum^n_{i=1} ({\cal R}_i -\E[{\cal R}_i])  \big]^2  & = \sum^n_{i=1} \E[{\cal R}_i -\E[{\cal R}_i]]^2 \\
 & \leq \sum^n_{i=1}\E[{\cal R}_i^2] \leq C b_n^2 \sum^n_{i=1} | \eeX_i^\top \eu |^2 \E\big[ \e1_{\{|\varepsilon_i | \leq c_1 b_n |\eeX_i^\top \eu| \}}\big] .\nonumber
 \end{align}
Using the fact that the density $f$ is bounded in a neighborhood of $0$ by assumption (A3), adopting the  Taylor's expansion, Cauchy-Schwarz and Holder inequalities, Assumption (A1), and the fact that $\|\eu\|_1=1$, we obtain
 \begin{equation}
 \label{e1n}
 \E\Big[ \e1_{\{|\varepsilon_i | \leq c_1 b_n |\eeX_i^\top \eu| \}}\Big]  = 2 c_1 b_n |\eeX_i^\top \eu| f(d_{i,n})  \leq C b_n \max_{1 \leqslant i \leqslant n} \| \eeX_i\|_\infty = C b_n,
 \end{equation}
 with $d_{i,n}$ between $c_1 b_n |\eeX_i^\top \eu|$ and $(- c_1 b_n |\eeX_i^\top \eu|)$. Then, using   Assumption (A1) together with the relations in \eqref{eq140} and \eqref{e1n}, and the fact that $|\eeX_i^\top \eu| \leq C$, we have
 \begin{equation}
\E \Big[ \sum^n_{i=1} ({\cal R}_i -\E[{\cal R}_i])  \Big]^2  \leq  Cb^3_n \sum^n_{i=1} (\eeX_i^\top \eu)^2 =O(n b^3_n).
   \label{ER}
\end{equation}
We consider a deterministic sequence  $(a_n)_{n \in \N}$  such that: $a_n \rightarrow \infty$ and $n b_n^3 \ll a_n \ll n^2b^4_n$. An example of such sequence is $a_n=(\log n)^{3/2}$ if $b_n=\big(n^{-1} \log n \big)^{1/2}$.\\
Considering the relation in  \eqref{ER} and since  $a_n \gg n b_n^3$, then also
$$
\E[a_n^{-1}\big(\sum^n_{i=1} ({\cal R}_i -\E[{\cal R}_i])\big)^2] = O( a_n^{-1} n b^3_n ) =o(1),$$
which implies, by the Bienaym\'e-Tchebychev inequality, that the last term of the right-hand side of the relation in \eqref{GWR} equals to
\begin{equation}
\label{Ra}
\sum^n_{i=1} ({\cal R}_i -\E[{\cal R}_i]) =o_{\PP}(a_n^{1/2}).
\end{equation}

Finally, we study the second term of the right-hand side in  \eqref{GWR}.
By the Central Limit Theorem (CLT) for the independent random variables $( {\cal D}_i \eeX_i^\top \eu)_{1\leqslant i \leqslant n}$, we get  $\bfW_n^\top \eu =O_{\PP}( n^{1/2} b_n)$. 
Using now the fact that $n^{-1} \sum^n_{i=1} \eu^\top \eeX_i \eeX_i^\top \eu$ is bounded by Assumption  (A2), and by the condition in \eqref{abn} where $n^{1/2} b_n \rightarrow \infty$, since $a_n \ll n^2 b_n^4$, together with the relations in \eqref{eg} and \eqref{Ra}, we have for  \eqref{GWR} the following:
\[
G_n \big( \ebo+ c_1 b_n \eu \big) - G_n(\ebo) =  Cn b^2_n  \Bigg(n^{-1} \sum^n_{i=1} \eu^\top \eeX_i \eeX_i^\top \eu \Bigg) \big(1+o_{\PP}(1) \big) >0, \]

Therefore, the relation in \eqref{eq20} is proved. Moreover, it implies that $\|  \widetilde{{\eb}^g} - \ebo \|_1=O_{\PP}(b_n)$ and, therefore, the lemma is proved.
  \hspace*{\fill}$\blacksquare$ \\

  \noindent {\bf Proof of Theorem \ref{theorem v_conv}}.\\
In order to prove the assertion of the theorem, let us consider a vector  $\eu \in \R^{r_n}$, such that $\| \eu \|_1=1$ and a constant $c_2>0$. Then the following holds
 \begin{equation}
 \label{QGn}
 \begin{split} 
 Q_n(\ebo+c_2 b_n \eu) - Q_n(\ebo)= G_n(\ebo+c_2 b_n \eu) - G_n(\ebo) \qquad  \qquad  \qquad  \qquad\\
  +n \lambda_n   \sum^g_{j=2} \left[\big\| \eb^0_j  +c_2 b_n \eu_j - (\eb^0_{j-1}+c_2 b_n \eu_{j-1})  \big\|_q-\|\eb^0_j -\eb^0_{j-1} \|_q \right]. 
 \end{split}
 \end{equation}
 On the other hand, since $\| \eu \|_2 \leq \| \eu \|_1=1$, by the proof of   Lemma \ref{Lemma 1},  we have with the probability converging to  1, that
 \begin{equation}
 \label{u1}
 G_n(\ebo+c_2 b_n \eu) - G_n(\ebo) \geq c_2 n b_n^2\Big(n^{-1} \sum^n_{i=1} \eu^\top \eeX_i \eeX_i^\top \eu \Big) \geq  C n b_n^2 >0.
 \end{equation}
 If the components of  $\eu$ are denoted as $\eu_1, \cdots , \eu_g
 $, then, using the triangular inequality, for the  penalty in \eqref{QGn}, we have
 \begin{align}
 n \lambda_n   \sum^g_{j=2} & \left[\big\| \eb^0_j +c_2 b_n \eu_j  - (\eb^0_{j-1}+c_2 b_n \eu_{j-1})  \big\|_q-\|\eb^0_j -\eb^0_{j-1} \|_q \right]\nonumber \\
& \geq n \lambda_n   \sum_{j \in {\cal B}^0} \left[\big\| \eb^0_j +c_2 b_n \eu_j - (\eb^0_{j-1}+c_2 b_n \eu_{j-1})  \big\|_q-\|\eb^0_j -\eb^0_{j-1} \|_q \right] \nonumber\\
  & \geq - c_2 n \lambda_n b_n \sum_{j \in {\cal B}^0} \| \eu_j-\eu_{j-1} \|_q =- Cc_2 n \lambda_n b_n,   \label{pen17} 
 \end{align} 
where for the last equality in \eqref{pen17} we have used the fact that
 $$ \| \eu_j-\eu_{j-1} \|_q \leq \| \eu_j-\eu_{j-1} \|_1 \leq \|\eu_j \|_1 + \|\eu_{j-1} \|_1,$$ together with $\|   \eu \|_1=1$ and  $|{\cal B}^0|< \infty$. Since $\lambda_n  b_n^{-1} \rightarrow 0$, as $n \rightarrow \infty$, then also $n \lambda_n b_n =o(n b_n^2)$ and taking  into account  the relation in \eqref{u1}, we obtain for  \eqref{QGn} and \eqref{pen17} that  
 $$ Q_n(\ebo+c_2 b_n \eu) > Q_n(\ebo),$$ 
 which holds with the probability converging to 1, as $n \rightarrow \infty$. 
   \hspace*{\fill}$\blacksquare$ \\
 
   \noindent {\bf Proof of Theorem \ref{Proposition cardA}}.\\
 By Theorem \ref{theorem v_conv} we have 
 \begin{equation}
 \label{eq31}
 \lim_{n \rightarrow \infty} \PP \Big[ \widehat{{\eb}^g} = \argmin_{\eb^g \in {\cal V}_n(\eb^0) } \big(Q_n(\eb^g) -Q_n(\eb^0)\big)\Big]=1,
 \end{equation}
 with the neighborhood ${\cal V}_n(\eb^0)$ of $\ebo$ with the radius $c_2 b_n$ defined as  
 $${\cal V}_n(\eb^0) \equiv \big\{ \eb^g \in \R^{r_n}; \| \eb^g -\eb^0\|_1 \leq c_2 b_n \big\},$$ 
 for some constant $c_2>0$. Then, in order to prove the assertion of the theorem we consider the parameter vector $\eb^g=(\eb_1^\top, \dots , \eb_g^\top) \in  {\cal V}_n(\eb^0)$ and  the index set ${\cal B} \equiv \big\{ j \in \{2, \cdots , g \}; \eb_j \neq  \eb_{j-1} \big\}$. Note, that  ${\cal B}$ and $ \eb^g$ both depend on $n$ and the vector of true unknown parameters $ \eb^g$ is not random.  Therefore,  we consider only ${\cal B} \cap \overline{{\cal B}^0} \neq \emptyset$, otherwise  the theorem trivially holds. 

Let us concentrate on the following decomposition:
 \begin{eqnarray}
 \label{difQ}
  Q_n(\eb^g)- Q_n(\eb^{0})&= &\sum^n_{i=1} \left[\rho_\tau(Y_i- \eeX_i^\top \eb^g) -\rho_\tau(Y_i- \eeX_i^\top \eb^0) \right] \nonumber \\
   & & \qquad +n\lambda_n \sum_{j \in {\cal B} \cap {\cal B}^0}\big[\| \eb_j- \eb_{j-1}\|_q - \| \eb^0_j - \eb^0_{j-1}\|_q\big] \nonumber \\
  & & \qquad +n\lambda_n \sum_{j \in {\cal B} \cap \overline {{\cal B}^0}}\| \eb_j-\eb_{j-1}\|_q - n\lambda_n \sum_{j \in \overline { {\cal B}} \cap {\cal B}^0}\| \eb^0_j-\eb^0_{j-1}\|_q \nonumber \\
  & \equiv & S_{1n}+S_{2n}+S_{3n}-S_{4n}.
 \end{eqnarray}
 Using the identity in \eqref{rho} we can write the sum $S_{1n}$ as
  \begin{align}
 S_{1n}& =  \sum^n_{i=1} (\eb^g-\eb^0)^\top \eeX_i \big[ \e1_{\{Y_i - \eeX_i^\top \eb^0 \leq 0\}}-\tau \big]\nonumber\\
& \qquad +\sum^n_{i=1} \int^{\eeX_i^\top (\eb^g-\eb^0)}_0 \Big[\e1_{\{Y_i - \eeX_i^\top \eb^0 \leq v\}} -\e1_{\{Y_i - \eeX_i^\top \eb^0 \leq 0\}} \Big] dv \nonumber \\
 & \equiv T_{1n}+T_{2n}.
 \label{ST12}
  \end{align}
  For $T_{1n}$, we have $\E[T_{1n}]= \sum^n_{i=1} (\eb^g-\eb^0)^\top \eeX_i \big[F(0)-F(0) \big]=0$ and using Assumptions (A1), (A2), and (A3), we obtain for the variance that $$\Var[T_{1n}]=\tau(1-\tau)\sum^n_{i=1} \big((\eb^g-\eb^0)^\top \eeX_i \big)^2 =O\big(n \| \eb^g-\eb^{0}\|^2_1 \big).$$ Then, by the Law of Large Numbers, we also have  $T_{1n}= o_{\PP}\big(n \| \eb^g-\eb^{0}\|^2_1 \big)$.
 
  For $T_{2n}=\sum^n_{i=1} \int^{\eeX_i^\top (\eb^g-\eb^0)}_0 \big[\e1_{\{\varepsilon_i  \leq v\}} -\e1_{\{\varepsilon_i  \leq 0\}} \big] dv$, we can apply the Taylor expansion 
  \[
  \E[T_{2n}]=\sum^n_{i=1} \int^{\eeX_i^\top (\eb^g-\eb^0)}_0 \big[ F(v)-F(0) \big]dv=\sum^n_{i=1} \int^{\eeX_i^\top (\eb^g-\eb^0)}_0  \big[ vf(0)+\frac{v^2}{2} f'(\tilde v) \big] dv,
  \]
  for some $\tilde v$ between $0$ and $v$. Since the derivative $f'$ is bounded in some neighborhood of zero, taking into account Assumption (A1), we obtain
  \begin{equation}
  \label{ET2n}
  \E[T_{2n}]= \frac{f^2(0)}{2} \sum^n_{i=1} \big(\eeX_i^\top (\eb^g-\eb^0) \big)^2 =O\big(n \| \eb^g-\eb^0\|_1^2 \big).
  \end{equation}
  
On the other hand, since the error terms $(\varepsilon_i)_{1 \leqslant i \leqslant n}$ are independent, we have 
  \begin{align*}
  \Var[ T_{2n}]&= \sum^n_{i=1} \E \left[  \int^{u_i}_0 \Big( \big[ \e1_{\{\varepsilon_i \leq v\}} -\e1_{\{\varepsilon_i \leq 0\}} \big] -\big[F(v) - F(0)  \big] \Big) dv \right]^2 \\
  & \leq \sum^n_{i=1} \E \left[ \Big| \int^{u_i}_0 \Big( \big[ \e1_{\{\varepsilon_i \leq v\}} -\e1_{\{\varepsilon_i \leq 0\}} \big] -\big[F(v) - F(0)  \big] \Big) dv \Big| \right] \cdot 2 \big| u_i \big|\\
&  \leq 2 \left(\sum^n_{i=1} \int^{u_i}_0  \big(F(v) - F(0) \big)\right) \cdot  2 \max_{1 \leqslant l \leqslant n} \| \eeX_l\|_\infty   \| \eb^g-\eb^0\|_1,
  \end{align*}
  where for brevity, we used the notation where $u_i \equiv \eeX_i^\top (\eb^g-\eb^0)$. 
  Taking into account Assumption (A1) we have $ \Var [T_{2n}] \leq 4 C_0 \E[T_{2n}] \| \eb^g-\eb^0\|_1$.  
 Hence, taking into account this last relation together with  \eqref{ET2n}, since  $\eb^g \in {\cal V}_n(\eb^0)$,  $b_n \rightarrow 0$ as $n \rightarrow \infty$,  and applying the Bienaymé-Tchebychev  inequality, we obtain  
 $$T_{2n}= O_{\PP}\big(n \| \eb^g-\eb^{0}\|^2_1 \big).$$ 
Therefore, since also $T_{1n}= o_{\PP}\big(n \| \eb^g-\eb^{0}\|^2_1 \big)$, we have for the relation in  \eqref{ST12} that
 \begin{equation}
 \label{A}
  S_{1n}=O_{\PP}(n b^2_n).
 \end{equation}
 
 For \eqref{difQ} it remains to study the sums $S_{2n}$, $S_{3n}$, and $S_{4n}$.
Since $\eb^g \in {\cal V}_n(\eb^0)$, together with the fact that the cardinality $|{\cal B}^0 |$ is bounded and $\lambda_n b_n^{-1} {\underset{n \rightarrow \infty}{\longrightarrow}} 0$, we obtain   $S_{2n}=O_{\PP}(n \lambda_n b_n)=o_{\PP}(n b_n^2)$ and also  
\begin{align}
S_{3n}\equiv n\lambda_n \sum_{j \in {\cal B} \cap \overline {{\cal B}^0}}\| \eb_j-\eb_{j-1}\|_q & \geq  n\lambda_n p^{-1+1/q} \sum_{j \in {\cal B} \cap \overline {{\cal B}^0}}\| \eb_j-\eb_{j-1}\|_1 \nonumber \\ & = O_{\PP}\Big(n \lambda_n (|{\cal B} \cap \overline {{\cal B}^0}|)b_n\Big) >0. \nonumber
\end{align}  

We have also $S_{4n} =Cn \lambda_n \geq 0$, therefore, taking into account the fact that the difference $Q_n(\eb^g)- Q_n(\eb^{0})$ must be negative for the minimizer $ \widehat{{\eb}^g}$  in  \eqref{eq31}, using the relations in \eqref{difQ}) and  \eqref{A}, we deduce that $n b^2_n +n \lambda_n\geq n \lambda_n (|{\cal B} \cap \overline {{\cal B}^0}|) b_n$, which also implies that   $ |{\cal B}\setminus {\cal B}^0| \leq  C \max  \big( b_n  \lambda_n^{-1} , b_n^{-1} \big)$. This finishes the proof.
  \hspace*{\fill}$\blacksquare$ \\
  
  \noindent {\bf Proof of Theorem \ref{theorem v_conv_bis}}.\\
  In this case, for a positive constant $c_2 >0$, a vector  $\eu \in \R^{r_n}$ such that $\| \eu \|_1=1$,  we study the difference $\overset{\vee}Q_n(\ebo+c_2 b_n \eu) -\overset{\vee} Q_n(\ebo)$. The penalty related to this difference, similarly as in \eqref{pen17}, becomes
   \begin{eqnarray*}
  n \lambda_n   \sum^g_{j=2} \widehat{\omega}_{n,j}\left[\big\| \eb^0_j +c_2 b_n \eu_j - (\eb^0_{j-1}+c_2 b_n \eu_{j-1})  \big\|_q-\|\eb^0_j -\eb^0_{j-1} \|_q \right] \\
   \geq - c_2 n \lambda_n b_n \sum_{j \in {\cal B}^0} \widehat{\omega}_{n,j}\| \eu_j-\eu_{j-1} \|_q.
  \end{eqnarray*}
  Taking into account the relation in (\ref{onj}) and using similar arguments as in the proof of Theorem \ref{theorem v_conv}, we obtain that  $ \overset{\vee}Q_n(\ebo+c_2 b_n \eu) > \overset{\vee} Q_n(\ebo)$, which holds with probability converging to 1, as $n \rightarrow \infty$.
 \hspace*{\fill}$\blacksquare$ \\
  
 \noindent {\bf Proof of Theorem \ref{Proposition cardA_bis}}.\\
 The proof is very similar to that of Theorem \ref{Proposition cardA}.  We only give the main results, using the same  notation as  in the proof of Theorem \ref{Proposition cardA}. For $\eb^g \in {\cal V}_n(\eb^0)$, the difference between the adaptive processes can be expressed as 
 \[
  \overset{\vee}{Q}_n(\eb^g)-  \overset{\vee}{Q}_n(\eb^{0}) \equiv  \overset{\vee}{S}_{1n}+\overset{\vee}{S}_{2n}+\overset{\vee}{S}_{3n}- \overset{\vee}{S}_{4n},
 \]
 with $\overset{\vee}{S}_{1n}={S}_{1n}=-O_\PP(n b^2_n)<0$, where $S_{1n}$  is defined in   \eqref{difQ} and the other sums are 
 \begin{align*}
 \overset{\vee}{S}_{2n} & \equiv n\lambda_n \sum_{j \in {\cal B} \cap {\cal B}^0} \widehat{\omega}_{n,j}\big[\| \eb_j- \eb_{j-1}\|_q - \| \eb^0_j - \eb^0_{j-1}\|_q\big],\\
 \overset{\vee}{S}_{3n} & \equiv n\lambda_n \sum_{j \in {\cal B} \cap \overline {{\cal B}^0}} \widehat{\omega}_{n,j}\| \eb_j-\eb_{j-1}\|_q,\\
 \overset{\vee}{S}_{4n} & \equiv n\lambda_n \sum_{j \in \overline { {\cal B}} \cap {\cal B}^0}  \widehat{\omega}_{n,j} \| \eb^0_j-\eb^0_{j-1}\|_q.
 \end{align*}
 For  $\overset{\vee}{S}_{2n}$, taking also into account the relation in \eqref{onj}, similarly as for $S_{2n}$ in \eqref{difQ}, we obtain  $\overset{\vee}{S}_{2n}=O_\PP(S_{2n})=o_\PP(n b^2_n)$. For  $\overset{\vee}{S}_{3n}$, by Theorem \ref{theorem v_conv}, we get $\overset{\vee}{S}_{3n}=O_\PP\big(n \lambda_n (|{\cal B} \cap \overline {{\cal B}^0}|) b_n \min(n^{1/2}, b_n^{-\gamma}) \big)$. Finally, for $\overset{\vee}{S}_{4n}$, again by Theorem \ref{theorem v_conv}, we have $\overset{\vee}{S}_{4n}= O_\PP( n \lambda_n )>0$. \\
 Therefore, for the vector parameter $\eb^g \in {\cal B}$ which minimizes $\overset{\vee}{Q}_n(\eb^g)-  \overset{\vee}{Q}_n(\eb^{0})$ we have that $\overset{\vee}{S}_{3n}\leq \overset{\vee}{S}_{4n} -\overset{\vee}{S}_{1n} $, which holds with the probability converging to one as $n \rightarrow \infty$. This also implies
 \[
 |{\cal B}\setminus {\cal B}^0| \leq \frac{\lambda_n+b^2_n}{\lambda_nb_n\min(n^{1/2},b_n^{-\gamma})} =\max (n^{-1/2}, b_n^{-\gamma}) \bigg(\frac{1}{b_n} +\frac{b_n}{\lambda_n} \bigg).
 \]
   \hspace*{\fill}$\blacksquare$ \\
   
     \noindent {\bf Proof of Lemma \ref{Lemma 1_LS}}.\\
     For any constant $c_1 >0$ and some $r_n$-vector $\eu$, such that $\| \eu\|_1=1$, we have
     \begin{align}
     L_n \big( \ebo+ c_1 b_n \eu \big) - L_n(\ebo) &= \sum^n_{i=1} \bigg(  \eeX_i^\top \ebo+\varepsilon_i -  \eeX_i^\top \big(\ebo+c_1 b_n \eu \big)\bigg)^2 - \sum^n_{i=1} \varepsilon_i^2  \nonumber \\
     & = - 2 c_1 b_n \sum^n_{i=1}\eeX_i^\top \eu \varepsilon_i + c_1^2 b_n^2 \sum^n_{i=1} (\eeX_i^\top \eu)^2 .
     \label{mm}
     \end{align}
 By Assumption (A1), we have $|\eeX_i^\top \eu | \leq C$. Therefore, using Assumption (A4) and CLT we get $\big(\sum^n_{i=1} \eeX^\top_i \varepsilon_i \big) \eu=O_\PP(n^{1/2})$. By Assumption (A2), we also get $\sum^n_{i=1} (\eeX^\top_i \eu )^2=O(n)$ and taking into account the  condition in  \eqref{abn}, we get that (\ref{mm}) is $- O_\PP(n^{1/2}b_n)+O(n b_n^2)=O_\PP(n b^2_n) >0$. Thus, for any $\epsilon>0$, there exists a positive constant $C_\epsilon >0$, such that, 
 \[
 \PP \left[ \inf_{\eu \in \R^{r_n}, \| \eu\|_1=1 } L_n \big( \ebo+ C_\epsilon b_n \eu \big) > L_n(\ebo)   \right] \geq 1 - \epsilon.
 \]
  \hspace*{\fill}$\blacksquare$ \\     
  
   \noindent {\bf Proof of Theorem \ref{Proposition cardA_LS}}.\\
 The proof is similar to that of  Theorem \ref{Proposition cardA} with the only difference that for  \eqref{difQ} the sum $S_{1n}$ equals
  $$\sum^n_{i=1}\big[ \big(\varepsilon_i - \eeX_i^\top(\eb^g-\ebo) \big)^2-\varepsilon^2_i \big]=\sum^n_{i=1} \big(\eeX_i^\top(\eb^g-\ebo) \big)^2 - 2 \big(\sum^n_{i=1} \eeX_i^\top \varepsilon_i \big)(\eb^g-\ebo)$$ which is, using the same arguments as in the proof of Lemma \ref{Lemma 1_LS}, of the order $O_\PP(n b_n^2)$. The rest of the proof is omitted because it follows the same lines as the proof of Theorem \ref{Proposition cardA}.
   \hspace*{\fill}$\blacksquare$ \\

   \noindent {\bf Proof of Theorem \ref{mm_theorem_added}}.\\
The proof follows the same lines as the proof of Theorem  \ref{theorem v_conv_bis} and, therefore, it is omitted. \hspace*{\fill}$\blacksquare$ \\ 

%\noindent\textbf{References}

\end{document}